\documentclass[aps,prd,twocolumn,a4paper,superscriptaddress,nofootinbib]{revtex4-1}
\usepackage{graphicx,caption,subcaption,amsmath,amsfonts,amssymb,multirow,extarrows,bm,acronym,float}
\usepackage[colorlinks,linkcolor=blue,citecolor=blue,urlcolor=blue ]{hyperref}
\floatstyle{plaintop}
\restylefloat{table}
\newcommand{\fs}{\mathcal{F}\text{-statistic}}

\newcommand{\Msun}{\,{\rm M}_\odot}

\newcommand{\DLUT}{School of Physics, Dalian University of Technology, Liaoning 116024, People's Republic of China}

\def\prd{Phys.~Rev.~D}       
\def\prl{Phys.~Rev.~Lett}    

\begin{document}

\title{Detection of a Higher Harmonic Quasi-normal Mode in the Ringdown Signal of GW231123}

\author{Hai-Tian Wang}
\email{wanght9@dlut.edu.cn}
\affiliation{\DLUT}
\author{Shao-Peng Tang}
\affiliation{Key Laboratory of Dark Matter and Space Astronomy, Purple Mountain Observatory, Chinese Academy of Sciences, Nanjing 210023, People's Republic of China}
\author{Peng-Cheng Li}
\affiliation{School of Physics and Optoelectronics, South China University of Technology, Guangzhou 510641, People’s Republic of China}
\author{Yi-Zhong Fan}
\email{yzfan@pmo.ac.cn}
\affiliation{Key Laboratory of Dark Matter and Space Astronomy, Purple Mountain Observatory, Chinese Academy of Sciences, Nanjing 210023, People's Republic of China}
\affiliation{School of Astronomy and Space Science, University of Science and Technology of China, Hefei, Anhui 230026, People's Republic of China}

\date{\today}

\begin{abstract}
The ringdown phase of a gravitational wave signal from a binary black hole merger offers a unique laboratory for testing general relativity in the strong-field regime and probing the properties of the final remnant black hole. In this study, we analyze the ringdown of GW231123 and find strong evidence for a multimode quasinormal spectrum. Our analysis employs two time-domain methodologies: a full Bayesian inference and an enhanced $\fs$ framework, which we extend to enable the calculation of Bayesian evidence and the reconstruction of posterior distributions for all model parameters. We report a statistically significant detection of the $\ell|m|n=200$ mode, with a $\log_{10}$(Bayes factor) of $5.3$, commencing at $12\,M$ after the peak amplitude--a time well within the accepted linear regime. This two-mode analysis yields a redshifted final mass of $305.6^{+35.7}_{-47.3}\Msun$ and a final spin of $0.84^{+0.07}_{-0.14}$ at $90\%$ credibility, from a ringdown signal with a network signal-to-noise ratio of approximately $14.5$. Furthermore, a test of the no-hair theorem performed using the two detected modes reveals no deviation from the predictions of general relativity. These results highlight the power of the $\fs$ methodology to uncover nuanced features in gravitational wave signals, thereby providing novel insights into the fundamental properties of black holes.
\end{abstract}

\maketitle

\acrodef{GW}{gravitational wave}
\acrodef{LIGO}{Laser Interferometer Gravitational-Wave Observatory}
\acrodef{LVKC}{LIGO-Virgo-KAGRA Collaboration}
\acrodef{LVC}{LIGO-Virgo Collaboration}
\acrodef{NR}{numerical relativity}
\acrodef{FD}{frequency-domain}
\acrodef{TD}{time-domain}
\acrodef{BH}{black hole}
\acrodef{TTD}{traditional time-domain}
\acrodef{BBH}{binary black hole}
\acrodef{GR}{general relativity}
\acrodef{PN}{post-Newtonian}
\acrodef{SNR}{signal-to-noise ratio}
\acrodef{PSD}{power spectral density}
\acrodef{PDF}{probability density function}
\acrodef{ACF}{auto-covariance function}
\acrodef{IMR}{inspiral-merger-ringdown}
\acrodef{QNM}{quasinormal mode}

\section{Introduction}\label{sec:intro}

Following the violent collision of a \ac{BBH}, the resulting remnant \ac{BH} oscillates and emits \acp{GW} as it settles into equilibrium \citep{Schw_PRD_Vishveshwara1970, GW_APJL_Press1971, QNM_APJ_Teukolsky1973}. This phase of the signal, known as the ``ringdown", can be mathematically described as a superposition of \acp{QNM} \citep{Berti:2009kk}. Each \ac{QNM} is a damped sinusoid characterized by spin-weighted spheroidal harmonics with angular indices $(\ell,m)$ and an overtone index $n$. The fundamental mode, $\ell |m| n = 2 2 0$, typically constitutes the dominant component of the ringdown emission.

According to the no-hair theorem, the entire spectrum of these modes is uniquely determined by the mass and spin of the remnant under the Kerr assumption \citep{Hawking:1971vc, PhysRevLett.34.905, kerr1963gw}. The ringdown signal, originating from the strong-field gravity regime, therefore offers a unique opportunity to test this fundamental tenet of \ac{GR}. Such a test requires the confident detection of at least two distinct \acp{QNM}, allowing for a consistency check of the remnant's properties \citep{Cardoso:2019rvt,Berti:2025hly}.

Beyond the fundamental mode, the ringdown signal may contain contributions from overtone modes ($n\geq 1$) and sub-dominant modes ($\ell|m|\neq 22$) \citep{Sperhake:2007gu}. However, robustly identifying these fainter modes is challenging due to the limited \ac{SNR} of the ringdown phase and the complexities of parameter estimation. 
For instance, an early claim of an overtone detection \citep{2019PhRvL.123k1102I} was later contested after the evidence vanished in a re-analysis that adopted a much higher sampling rate \citep{2022PhRvL.129k1102C}. Subsequent studies demonstrated that this discrepancy arises from an improper downsampling method in the initial analysis, which can corrupt frequencies far below the Nyquist limit \citep{Wang:2023mst, Wang:2024yhb}. The effects of such downsampling choices have been studied in further detail \citep{Siegel:2024jqd}. Furthermore, multiple independent analyses have confirmed that there is only weak evidence for the first overtone in GW150914 \citep{Finch:2022ynt,Correia:2023bfn}.

A critical aspect of ringdown analysis is the choice of the starting time. Because overtone modes decay much more rapidly than the fundamental mode, there is a temptation to begin the analysis close to the signal's peak amplitude to capture them. However, this early post-merger region is dominated by non-linear dynamics where the linear \ac{QNM} framework is not physically applicable, implying that such fits could be unphysical. Indeed, several studies using numerical waveforms have argued that fitting for higher overtones $(n>2)$ at these early times can lead to overfitting transient radiation and non-linearities from the merger itself \citep{Baibhav:2023clw,Nee:2023osy,Zhu:2023mzv,Clarke:2024lwi}. The debate is further nuanced by recent work from \citet{Giesler:2024hcr}, which confirmed the existence of higher overtones and second-order modes in numerical waveforms and argued that some non-linear effects might still be describable within a linear perturbation framework. This highlights the ongoing need for careful investigation into the role of overtones in real \ac{GW} data analysis for \ac{BH} spectroscopy.
The search for sub-dominant modes faces similar challenges. The sole claim of a sub-dominant mode detection in GW190521 \citep{Capano:2021etf} was identified in the early, non-linear regime, approximately $6\,M$ after the peak, whereas a more conservative start time of at least $10\,M$ is often recommended for the linear perturbative description to be valid \citep{Mitman:2025hgy,Berti:2025hly}. Subsequent studies found only weak evidence for sub-dominant modes in this event \citep{Siegel:2023lxl,Gennari:2023gmx}.

To overcome these difficulties, advanced and efficient data analysis methods are essential. The $\fs$ method, originally formulated for continuous \ac{GW} signals \citep{Jaranowski:1998qm,Cutler:2005hc,Dreissigacker:2018afk,Sieniawska:2019qnx} and later applied to other sources like extreme mass-ratio inspirals \citep{Wang:2012xh}, has recently been adapted for ringdown analysis \citep{Wang:2024jlz, Wang:2024yhb}. By analytically maximizing over linear parameters such as the amplitudes and phases of the \acp{QNM}, the $\fs$ reduces the dimensionality of the parameter space, which can enhance the efficiency and robustness of the search. However, previous applications of this method lacked a formalism for reconstructing the posteriors of these maximized parameters or for computing the Bayesian evidence needed for model comparison. In this work, we resolve these two outstanding issues and apply the now complete framework to a new \ac{GW} event.

The recently detected event GW231123, originating from a high-mass, high-spin system, presents an ideal candidate for black hole spectroscopy \citep{LIGOScientific:2025rsn}. Its signal is dominated by a high-\ac{SNR} merger-ringdown phase, making it a prime target for identifying multimode signatures. 
Similar to \citet{LIGOScientific:2025rsn}, we find strong evidence for the $200$ mode, with an amplitude larger than that of the fundamental mode. 
Moreover, we report a detection of the $220+200$ combination at a time delay of $\Delta t=12\,M$ post-peak with the strongest evidence observed among all combinations.
Additionally, the inferred remnant parameters from this two-mode analysis are consistent with full \ac{IMR} analyses using both the \textbf{NRSurd7q4} \citep{Varma:2019csw} and \textbf{SEOBNRv5PHM} \citep{Ramos-Buades:2023ehm} waveform models, without exhibiting bimodal posterior structures. Based on this detection, we perform a no-hair theorem test and find no deviations from \ac{GR}. We also search for evidence of a third \ac{QNM} but do not find strong support for its presence. This analysis highlights the enhanced performance of the $\fs$ method, demonstrating its efficiency and reliability in uncovering the detailed features of ringdown signals.

\begin{figure}
\centering
\includegraphics[width=0.48\textwidth,height=5cm]{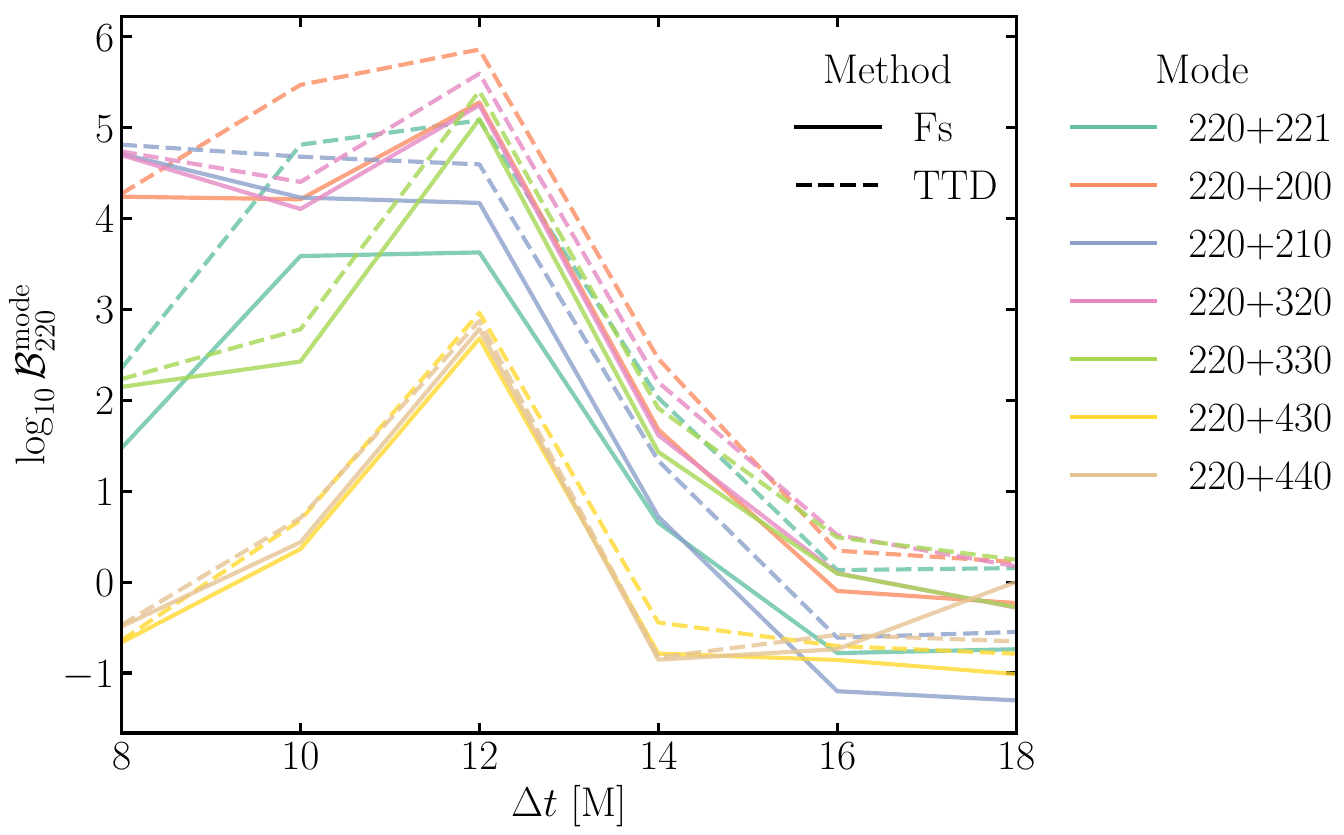}
\caption{
Bayesian evidence for various two-mode \ac{QNM} combinations in the ringdown analysis of GW231123. The vertical axis shows the log$_{10}$(Bayes factor) of each combination relative to a model containing only the fundamental ($\ell=|m|=2, n=0$, or $220$) mode. The horizontal axis represents the analysis start time ($\Delta t$) as a delay from the signal's polarization peak. Results from two independent analysis methods are compared: the $\fs$ method (solid lines, labeled ``Fs") and the \ac{TTD} method (dashed lines). Different colors correspond to different \ac{QNM} combinations as detailed in the legend. Both methods yield qualitatively consistent results, showing a clear preference for the $220+200$ combination, with the evidence peaking at a start time of $\Delta t = 12M$.
}\label{fig:bfs_m2}
\end{figure}


\section{Multimode search}\label{sec:bayes}
\begin{figure*}
\centering
\begin{subfigure}[b]{0.48\linewidth}
\centering
\includegraphics[width=\textwidth,height=8cm]{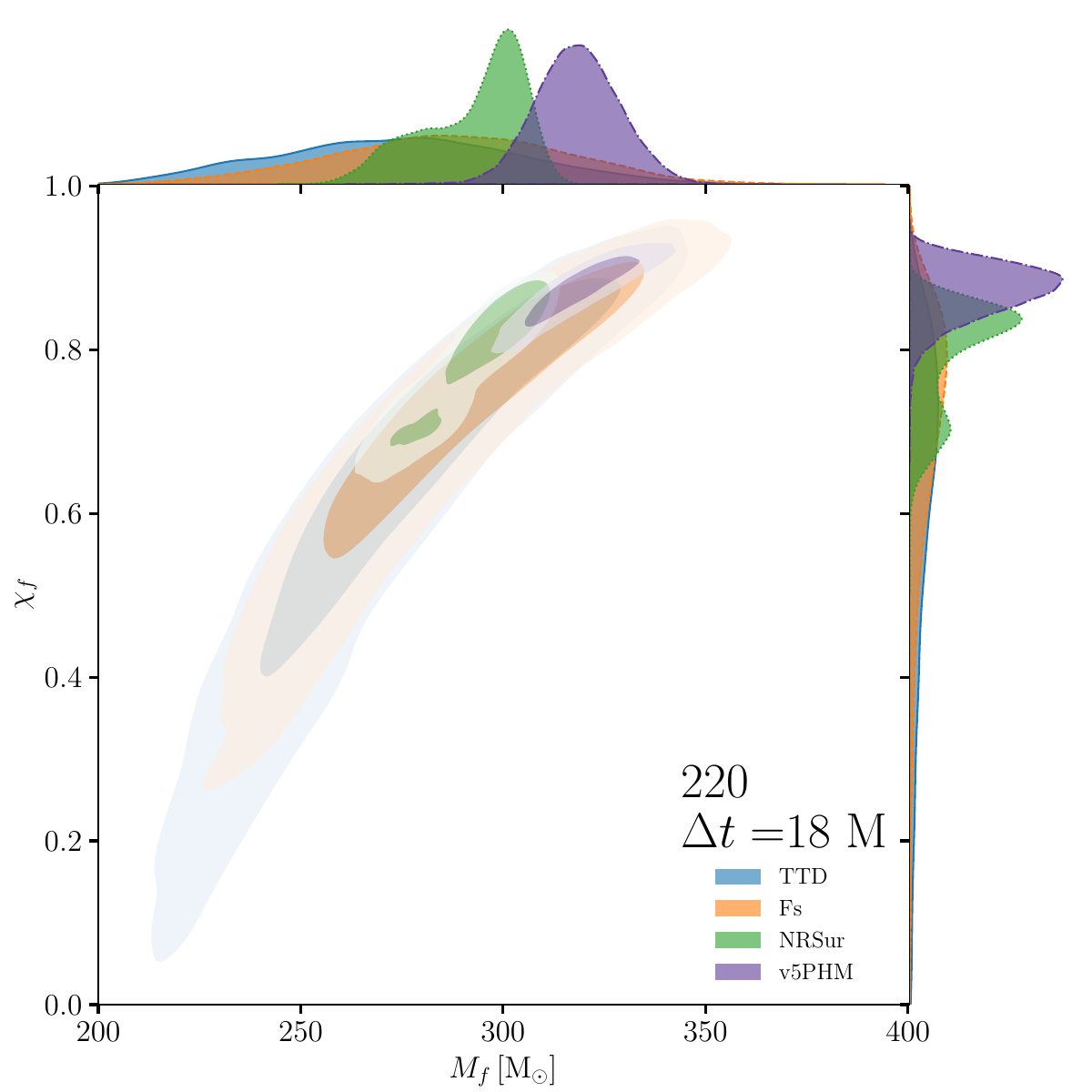}
\end{subfigure}%
\begin{subfigure}[b]{0.48\linewidth}
\centering
\includegraphics[width=\textwidth,height=8cm]{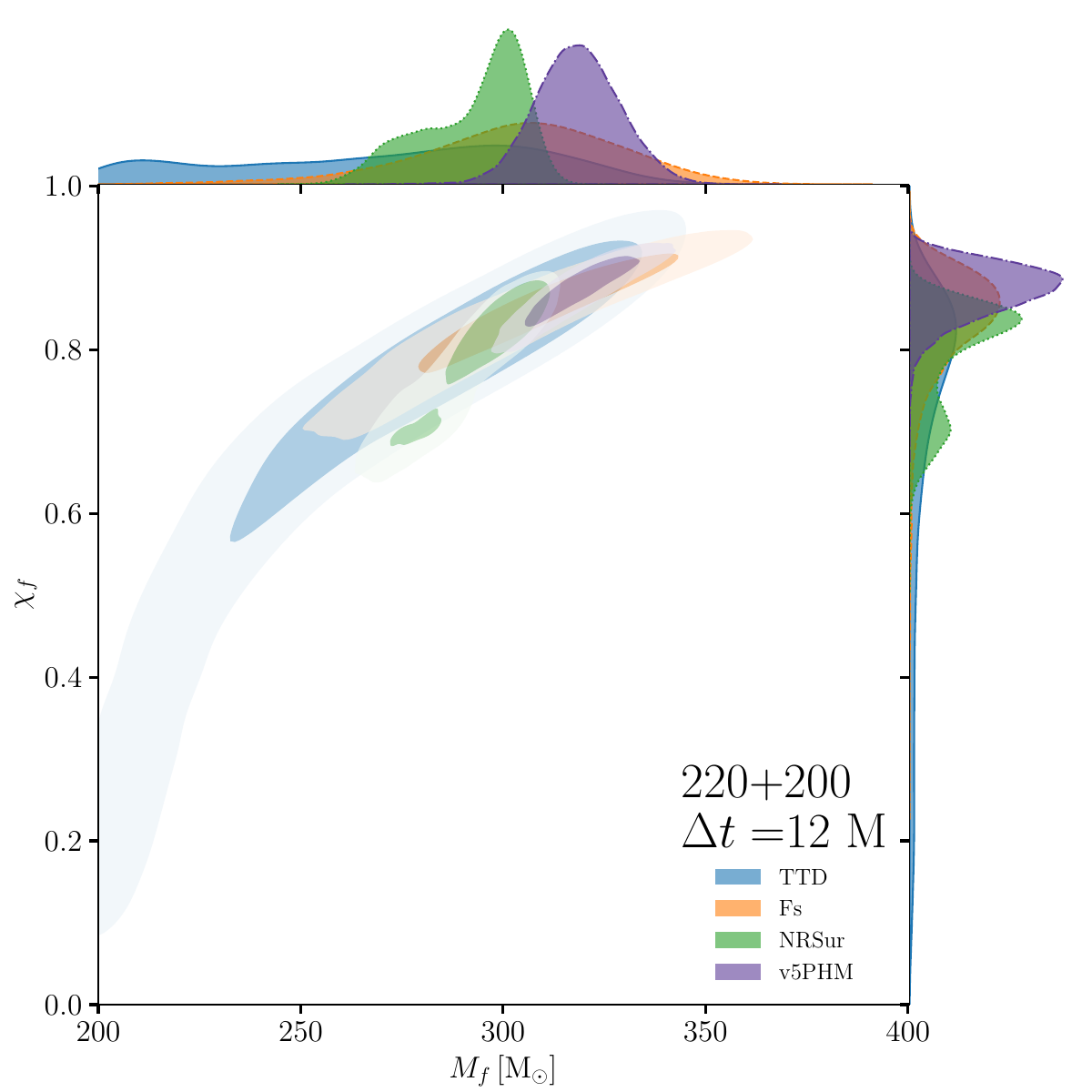}
\end{subfigure}%
\caption{
Posterior distributions for the redshifted final mass $M_f$ and final spin $\chi_f$ of the GW231123 remnant. The results from our ringdown analyses, using the \ac{TTD} method (blue) and the $\fs$ method (orange), are compared against posteriors from full \ac{IMR} analyses using the \textbf{NRSur7dq4} (green) and \textbf{SEOBNRv5PHM} (purple) waveform models.
Left panel: An analysis including only the fundamental mode, shown at a start time of $\Delta t=18\,M$.
Right panel: The analysis incorporating the sub-dominant $(\ell |m|n)=(200)$ mode, shown at its optimal start time of $\Delta t=12\,M$.
The contours enclose the $60\%$ and $90\%$ credible regions. The top and right side panels show the corresponding one-dimensional marginalized posterior distributions.
}\label{fig:fmfs_m2}
\end{figure*}

We perform a \ac{TD} ringdown analysis of GW231123 using two distinct methodologies. The first is the \ac{TTD} method \citep{Wang:2023mst,Wang:2024jlz, Wang:2024yhb}, which stochastically samples the full parameter space. The second is the $\fs$ method, which enhances efficiency by analytically maximizing the likelihood over the linear parameters (amplitudes and phases) of the \acp{QNM}. The efficiency and robustness of the $\fs$ for ringdown analyses have been established in previous work \citep{Wang:2024jlz,Wang:2024yhb}. For a consistent comparison, both methods utilize the same processed strain data and estimated \ac{ACF} in their likelihood calculations (see Sec.~\textbf{Appendix} \ref{sec:methods} for details).

To isolate the ringdown signal and mitigate contamination from the pre-merger phase, we fix the sky location and polarization angle $(\text{RA}, \text{DEC}, \psi)$ to their maximum likelihood values $(3.24, 0.25, 2.23)$ derived from a full \ac{IMR} analysis with the \textbf{NRSur7dq4} waveform model \citep{Varma:2019csw, LIGOScientific:2025rsn}. To account for the uncertainty in the onset of the linear ringdown phase, we vary the analysis start time, $\mathrm{t}_c$, in discrete steps of $2M$ over a range of $[8, 18]M$ after the signal's polarization peak, $t^{\mathrm{pol}}_{\mathrm{c}}$. For GW231123, the redshifted remnant mass is approximately $M=298\Msun$ \citep{LIGOScientific:2025rsn}, and the peak time is $t^{\mathrm{pol}}_{\mathrm{c}} = 1384782888.615373$ GPS.\footnote{In this context, $1M$ corresponds to approximately $1.5$ ms. For the fixed sky location, the polarization peak time at the LIGO Hanford detector is $1384782888.5998$ GPS.}

Our search focuses on identifying a sub-dominant mode accompanying the fundamental ($\ell|m|n=220$) mode, selecting candidates from the set $\ell|m|n \in \{221, 200, 210, 320, 330, 430, 440\}$. As shown in Fig.~\ref{fig:bfs_m2}, we compute the Bayes factor for each two-mode combination against the fundamental-mode-only model across the range of start times. Both the \ac{TTD} and $\fs$ methods overwhelmingly favor the $220+200$ combination, with the evidence peaking at a time delay of $\Delta t=12M$ post-peak. At this start time, the Bayes factors are as high as $10^{5.9}$ (\ac{TTD}) and $10^{5.3}$ ($\fs$), providing strong evidence for the presence of the $200$ mode within a region widely accepted to be governed by linear perturbation theory. The network \ac{SNR} for this two-mode signal is approximately $14.5$ at $\Delta t=12M$, decaying to $10.6$ by $\Delta t=20M$.

\begin{figure*}
\centering
\includegraphics[width=0.88\textwidth,height=12cm]{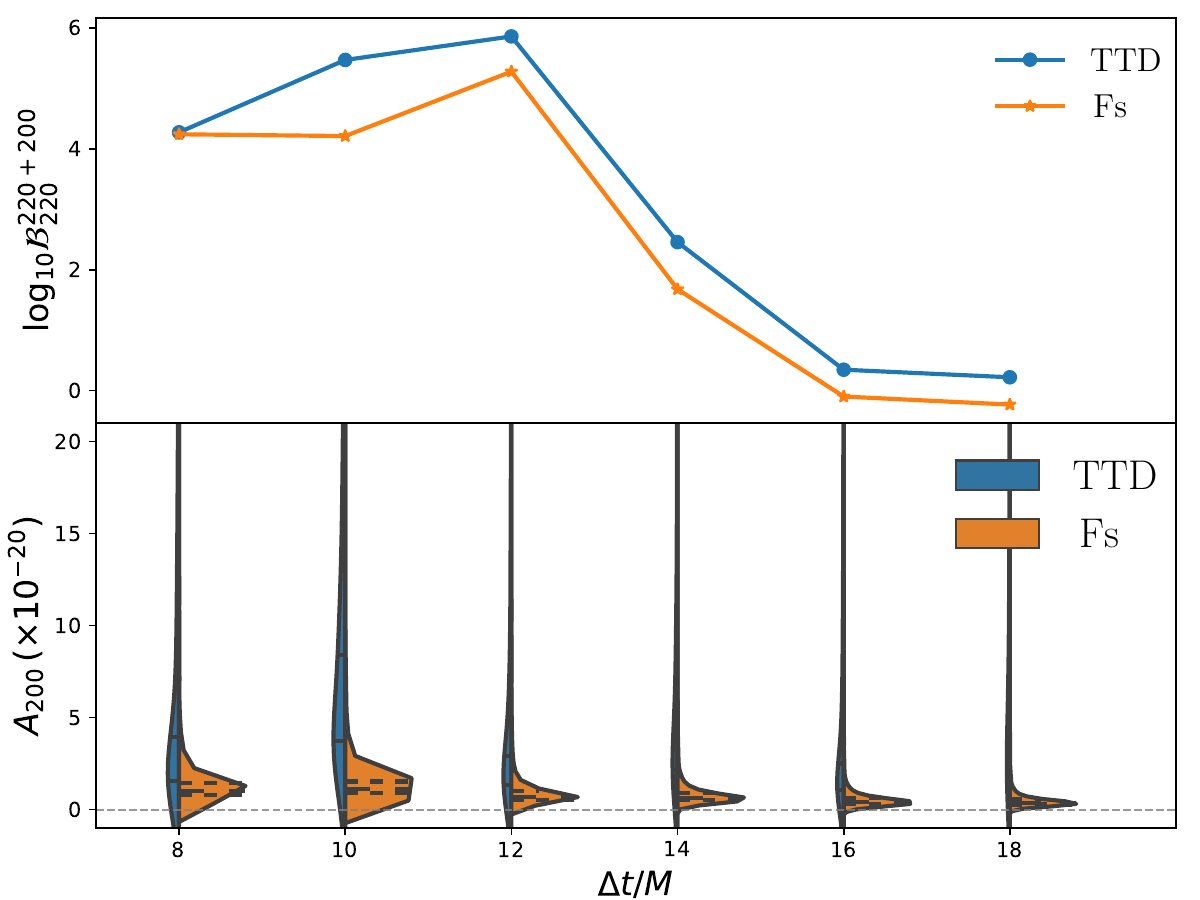}
\caption{
A direct comparison of the analysis of the $220+200$ mode combination using the $\fs$ (orange lines/right violins) and \ac{TTD} (blue lines/left violins) methods.
Top panel: The log$_{10}$(Bayes factor), $\log_{10} \mathcal{B}^{220+200}_{220}$, in favor of the two-mode model over the fundamental-mode-only model, shown as a function of the analysis start time $\Delta t$. The statistical evidence for including the $200$ mode peaks sharply at $\Delta t=12\,M$ and declines rapidly at later times.
Bottom panel: Violin plots showing the posterior distributions for the amplitude of the $200$ mode, $A_{200}$, at the corresponding start times. The peak in Bayesian evidence at $\Delta t=12\,M$ directly corresponds to a confidently measured, non-zero amplitude. Both panels show a high degree of consistency between the two analysis methods.
}\label{fig:bfs200}
\end{figure*}

Fig.~\ref{fig:fmfs_m2} demonstrates that the remnant parameters inferred from the $220+200$ model are remarkably consistent with the results of full \ac{IMR} analyses using both the \textbf{NRSur7dq4} \citep{LIGOScientific:2025rsn} and \textbf{SEOBNRv5PHM} \citep{Ramos-Buades:2023ehm} waveform models. It is noteworthy that while our analysis start time is defined using the \textbf{NRSur7dq4} maximum likelihood, our results also show significant overlap with the \textbf{SEOBNRv5PHM} posterior, which has a different peak time definition. This consistency underscores the robustness of our ringdown analysis across different IMR waveform models. 
In contrast, an analysis using only the fundamental mode fails to produce results consistent with the IMR posteriors until much later start times ($\Delta t \ge 16, 18\,M$), after which the \ac{SNR} is significantly lower and the constraints become less meaningful. For the favored $220+200$ model at $\Delta t=12\,M$, we constrain the redshifted final mass and spin to be $305.6^{+35.7}_{-47.3}\Msun$ and $0.84^{+0.07}_{-0.14}$ ($\fs$), compared to $276.8^{+47.1}_{-71.9}\Msun$ and $0.77^{+0.11}_{-0.55}$ (\ac{TTD}), at $90\%$ credible level.
As a quantitative comparison, for the $220$ mode starting from $18\,M$ after the peak, the constraints on the redshifted final mass and final spin given by the $\fs$ (TTD) method are $286.8^{+45.2}_{-48.4}\Msun$ ($274.5^{+49.3}_{-48.8}\Msun$) and $0.73^{+0.15}_{-0.37}$ ($0.66^{+0.20}_{-0.45}$) at $90\%$ credible level, respectively.

\begin{figure*}
\centering
\includegraphics[width=0.88\textwidth,height=12cm]{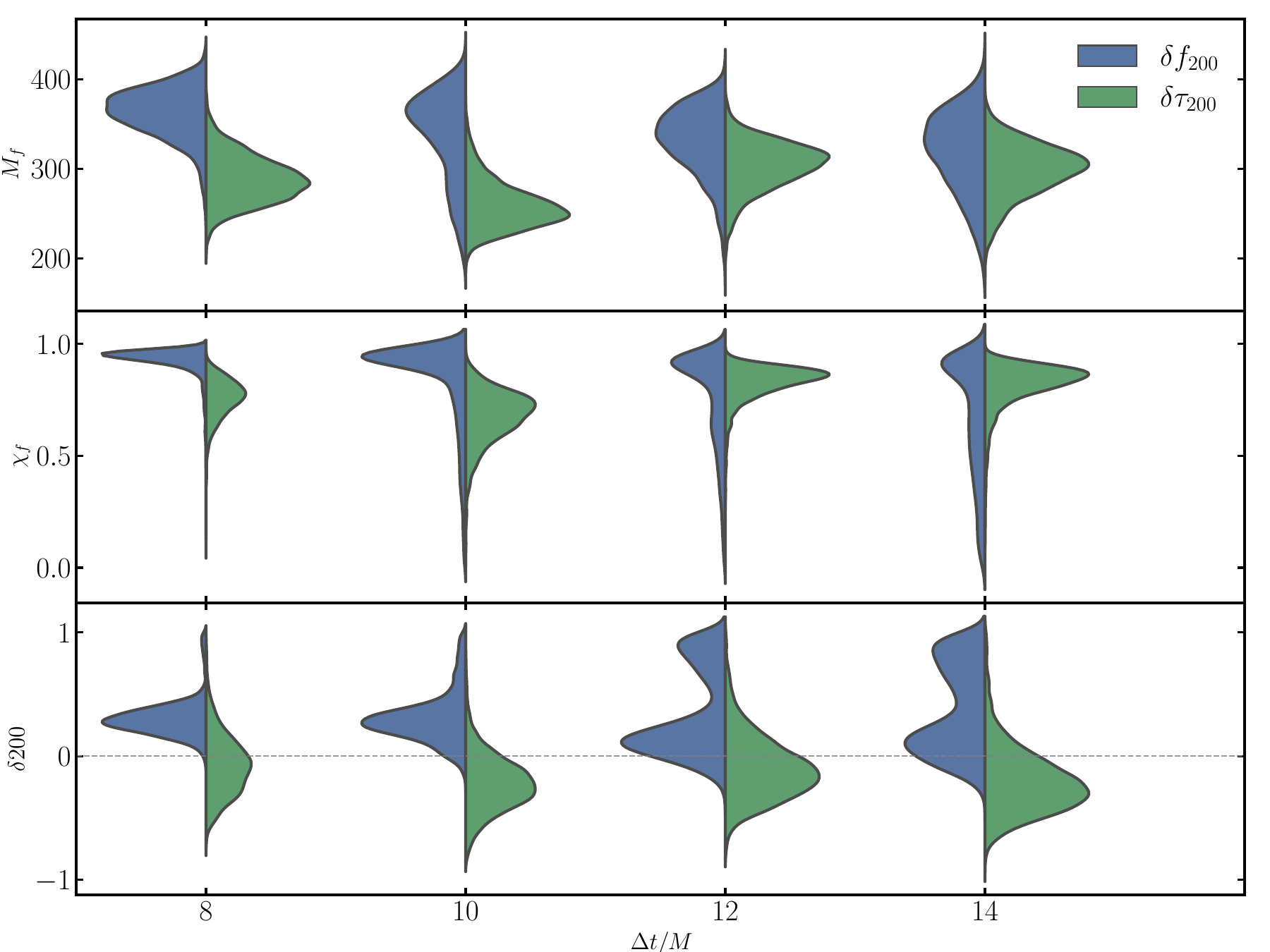}
\caption{
Tests of the no-hair theorem using the $220+200$ mode combination in the ringdown signal of GW231123. The x-axis represents the time interval $(\Delta t)$ between the start of the ringdown analysis and the strain peak. Each of the four $\Delta t$ values is represented by two violin plots (left and right columns). The left column shows results when allowing deviations in the frequency of the $200$ mode, \( f_{200} \times (1 + \delta f_{200}) \), with the bottom panel displaying the posterior distribution of \(\delta f_{200}\), the top panel showing the final mass \( M_f \), and the middle panel showing the final spin \( \chi_f \). The right column presents analogous results for deviations in the damping time of the $200$ mode, \( \tau_{200} \times (1 + \delta \tau_{200}) \), with the bottom panel showing \(\delta \tau_{200}\), the top panel \( M_f \), and the middle panel \( \chi_f \). The violin plots illustrate the posterior distributions across different analysis start times, providing constraints on potential deviations from general relativity predictions.
}\label{fig:dfdt200}
\end{figure*}

A direct comparison of the methodologies, presented in Fig.~\ref{fig:bfs200}, reveals that while the overall trend of the Bayes factor is consistent between the two methods, the $\fs$ yields significantly more stringent constraints on the linear parameters, such as the amplitude of the $200$ mode ($A_{200}$). Similarly, for the final mass and spin, the constraints derived from the $\fs$ are tighter than those from the \ac{TTD} analysis. This improved performance is a direct consequence of the reduced dimensionality of the parameter space sampled in the $\fs$ method, which allows for a more efficient exploration of the posterior, consistent with previous findings \citep{Wang:2024jlz}.

Finally, we find limited support for other two-mode combinations. Although the $220+221$ model produces a posterior that marginally overlaps with the full \ac{IMR} results, its Bayes factor is much lower than that of the $220+200$ model (e.g., $\log_{10}\mathcal{B}^{220+221}_{220+200}=-1.7$ at $\Delta t=12\,M$). The remaining combinations fail to produce physically meaningful posteriors that are consistent with the \ac{IMR} analysis. 
We also extend our search to investigate the presence of a third mode in the ringdown signal. Using the $\fs$ method, we analyze three-mode combinations consisting of the fundamental mode plus two additional modes chosen from the set $\{221, 200, 210, 320, 330\}$. We compute the Bayes factors for these combinations relative to the preferred $220+200$ model. The most favorable three-mode combination is $220+200+330$, which achieves a modest Bayes factor of $\log_{10}\mathcal{B}=1.3$ over the two-mode case, but only at a very early start time of $\Delta t=8\,M$. Further details are provided in Sec.~\textbf{Appendix}, and no other three-mode combination is favored over the $220+200$ model.
Given the marginal statistical support and the analysis's reliance on the early, likely non-linear regime, we conclude that there is no compelling evidence for a third \ac{QNM} in GW231123.

Building on the detection of the $220+200$ mode combination, we perform a test of the no-hair theorem using the $\fs$ method. We independently parameterize potential deviations in the oscillation frequency and damping time of the sub-dominant mode as $f_{200}(1+\delta f_{200})$ and $\tau_{200}(1+\delta \tau_{200})$, where $\delta f_{200}=0$ and $\delta \tau_{200}=0$ correspond to the predictions of \ac{GR}. As shown in Fig.~\ref{fig:dfdt200}, the constraints on $\delta \tau_{200}$ are consistent with zero across all analysis start times. In contrast, the analysis shows a deviation in $f_{200}$ when initiated at earlier times ($\Delta t=8,10\,M$), likely due to contamination from non-linear merger dynamics or the influence of other unaccounted-for modes. However, at our preferred start time of $\Delta t=12\,M$, the analysis reveals no evidence for deviations from \ac{GR}. The constraints at the $90\%$ credible level are $\delta f_{200}=0.26^{+0.68}_{-0.38}$ and $\delta \tau_{200}=-0.24^{+0.51}_{-0.32}$, both of which are consistent with zero.


\section{Discussion and Conclusions}\label{sec:con}
Our analysis of GW231123 provides strong evidence for a multimode ringdown signal. 
The key finding of this work is the statistically significant detection of the $\ell|m|n=200$ mode, which is supported by both the $\fs$ and \ac{TTD} methods at a start time of $\Delta t=12\,M$ post-peak, with $\log_{10}$ Bayes factors of $5.3$ and $5.9$ over the fundamental-mode-only hypothesis.
At the maximum likelihood, the amplitude of the $200$ mode is approximately an order of magnitude larger than that of the fundamental $220$ mode, indicating that this sub-dominant mode dominates the observed ringdown signal. 
Specifically, at the maximum likelihood, the amplitudes of the $220$ and $200$ modes at the start time of $12\,M$ are $1.4\times 10^{-21}$ and $2.9\times 10^{-20}$ for the $\fs$ method, and $3.1\times 10^{-21}$ and $3.9\times 10^{-20}$ for the \ac{TTD} method, respectively. 
This result is marginally consistent with the ringdown analysis presented in \citet{LIGOScientific:2025rsn}, note that their analysis starts at a later time and yields different best-fit amplitudes (approximately $6\times 10^{-22}$ for the $200$ mode vs. $2\times 10^{-22}$ for the $220$ mode at $\Delta t\sim 28\,M$).
Notably, considering the amplitude damping, the amplitude ratio $A_{200}/A_{220}$ inferred at $\Delta t=12\,M$  will also be significantly larger at $\Delta t=28\,M$ compared to the ratio obtained in \citet{LIGOScientific:2025rsn}. This should be due to the different prior probability densities and noise estimation methods used in \citet{LIGOScientific:2025rsn} and our work.

The pronounced dominance of the $200$ mode raises questions about its origin. One plausible explanation is that the source was a strongly precessing binary system. As shown by \citet{Zhu:2023fnf}, strong precession is known to amplify sub-dominant modes like the $200$ mode. The current generation of \ac{IMR} waveform models may struggle to accurately capture the effects of high spins, which could lead to an underestimation of the $200$ mode's contribution in a full \ac{IMR} analysis \citep{LIGOScientific:2025rsn}. 
Nearly axisymmetric plunge dynamics (e.g., high-eccentricity or head-on encounters) can also excite $200$ mode \citep{CalderonBustillo:2020xms,Cheung:2022rbm}. In addition, an edge-on inclination can suppress the observed $220$ emission, making $200$ appear comparatively stronger \citep{Rossello-Sastre:2025gtq}. 
Beyond these possibilities, more exotic scenarios remain: head-on collisions of boson stars have been shown to excite the $200$ mode \citep{Palenzuela:2006wp, CalderonBustillo:2020fyi}. Given the known degeneracy between the \ac{GW} signals from quasi-circular precessing mergers and those from extremely eccentric or head-on collisions \citep{CalderonBustillo:2020xms}, our ringdown analysis may also suggest that GW231123 is a viable candidate for a head-on collision of two boson stars or two highly spinning Proca stars.

Our investigation of a potential third mode, specifically the $220+200+330$ combination, yielded intriguing but inconclusive results. While this model showed a marginal preference at very early start times ($\Delta t=8\,M$ and $10\,M$) and produced posteriors consistent with the \textbf{NRSur7dq4} \ac{IMR} analysis (see Sec.~\textbf{Appendix} \ref{A3:3m}), its statistical evidence was weaker than that of the two-mode model at $\Delta t=12\,M$. Furthermore, analyses beginning at such early times are likely contaminated by non-linear dynamics. We therefore consider this three-mode search an exploratory attempt and refrain from drawing strong physical conclusions from it.

Methodologically, this work highlights the power and robustness of our analysis framework. The $\fs$ method demonstrated superior performance over the conventional \ac{TTD}, particularly in yielding tighter constraints for a comparable number of posterior samples, and was faster by at least an order of magnitude \citep{Wang:2024jlz}. This efficiency gain is a direct benefit of its reduced-dimensionality parameter space. Moreover, our ringdown analysis proved to be robust. Although the start time was determined using the maximum likelihood from the \textbf{NRSur7dq4} model, the resulting remnant parameters showed significant overlap with posteriors from both the \textbf{NRSur7dq4} and \textbf{SEOBNRv5PHM} models. This underscores the value of ringdown analysis as a largely independent and complementary probe of the final state of binary black hole mergers.

\begin{acknowledgments}
This work is supported in part by ``the Fundamental Research Funds for the Central Universities" at Dalian University of Technology and by National Natural Science Foundation of China under grants No. 12233011 and No. 12303056. 
This research has made use of data or software obtained from the Gravitational Wave Open Science Center~\cite{gwosc-url}, a service of LIGO Laboratory, the LIGO Scientific Collaboration, the Virgo Collaboration, and KAGRA~\cite{KAGRA:2023pio}. 
LIGO Laboratory and Advanced LIGO are funded by the United States National Science Foundation (NSF) as well as the Science and Technology Facilities Council (STFC) of the United Kingdom, the Max-Planck-Society (MPS), and the State of Niedersachsen/Germany for support of the construction of Advanced LIGO and construction and operation of the GEO600 detector. 
Additional support for Advanced LIGO was provided by the Australian Research Council. 
Virgo is funded, through the European Gravitational Observatory (EGO), by the French Centre National de Recherche Scientifique (CNRS), the Italian Istituto Nazionale di Fisica Nucleare (INFN) and the Dutch Nikhef, with contributions by institutions from Belgium, Germany, Greece, Hungary, Ireland, Japan, Monaco, Poland, Portugal, Spain.
KAGRA is supported by Ministry of Education, Culture, Sports, Science and Technology (MEXT), Japan Society for the Promotion of Science (JSPS) in Japan; National Research Foundation (NRF) and Ministry of Science and ICT (MSIT) in Korea; Academia Sinica (AS) and National Science and Technology Council (NSTC) in Taiwan of China.
\end{acknowledgments}


\bibliographystyle{apsrev4-1}
\bibliography{fs231123}

\appendix

\section{Methodology}\label{sec:methods}

\subsection{The Ringdown Waveform Model}
The gravitational wave emission during the ringdown phase is modeled as a superposition of quasinormal modes (\acp{QNM}), which represent the characteristic vibrations of the final, settled black hole. Each individual \ac{QNM} is identified by a set of indices: the angular numbers $(\ell, m)$ and the overtone number $n=0,1,2\ldots$. Physically, each mode is a damped sinusoid characterized by four parameters: an amplitude $A_{\ell mn}$, a phase $\phi_{\ell mn}$, an oscillation frequency $f_{\ell mn}$, and a damping time $\tau_{\ell mn}$. For convenience, the latter two are often combined into a single complex frequency for each mode, defined as $\Omega_{\ell mn} = 2\pi f_{\ell mn} + i/\tau_{\ell mn}$.

Assuming the remnant of GW231123 is a Kerr \ac{BH}\footnote{By default, the oscillation frequency $f_{\ell mn}$ and final mass $M_f$ are all defined in the detector frame.}, the no-hair theorem imposes a critical constraint: the frequencies and damping times of all \acp{QNM} are not independent parameters. Instead, they are uniquely determined by the remnant's final mass $M_f$ and dimensionless spin $\chi_f$. These values can be precisely computed using black hole perturbation theory \citep{Leaver:1985ax, Berti:2005ys, Berti:2009kk}.

The complete time-domain waveform, combining the contributions from all excited modes, is expressed through its polarization components, $h_+(t)$ and $h_\times(t)$, as follows:
\begin{equation}
\begin{aligned}
	&h_+(t) + i h_{\times}(t) \\
	&= \sum_{\ell,m,n}{}_{-2}S_{\ell m}(\iota, \delta; \chi_f)A_{\ell m n} \exp [ i (\Omega_{\ell m n} t + \phi_{\ell m n} )] .
\end{aligned}
\end{equation}
Here, the term ${}_{-2}S_{\ell m}(\iota, \delta; \chi_f)$ represents the spin-weighted spheroidal harmonics of spin weight $-2$. It describes the angular pattern of the emitted radiation, which depends on the inclination angle $\iota$ between the \ac{BH}'s spin and the observer's line of sight. The azimuth angle $\delta$ is degenerate with the individual mode phases and is therefore set to zero by convention.

Given the limited \ac{SNR} of the ringdown signal for GW231123, we employ a common simplification by approximating the spin-weighted spheroidal harmonics with their spherical counterparts, ${}_{-2}Y_{\ell m}(\iota, \delta)$. 
This approximation introduces a useful symmetry between modes with positive and negative $m$ indices, given by $h_{\ell m}=(-1)^{\ell}h^{*}_{\ell-m}$, where the asterisk denotes complex conjugation.

\subsection{The traditional time-domain method}

Our first analysis approach is the \ac{TTD} method. This method aims to infer the posterior distributions of all model parameters, denoted by the vector $\Theta$, by directly evaluating the likelihood function for the data. The observed data in a detector, $d(t)$, is modeled as the sum of the \ac{GW} signal $h(t,\Theta)$ and stationary, Gaussian noise $n(t)$. To isolate the ringdown, we use a truncated segment of the data, a technique well-established in previous studies \citep{Isi:2021iql,Wang:2023mst,Wang:2024liy,Siegel:2024jqd}.

For Gaussian noise, the \ac{TD} log-likelihood function, $\ln\mathcal{L}$, is given by:
\begin{equation}
\begin{aligned}\label{eq:ll0}
\ln\mathcal{L}(\Theta) &= -\frac{1}{2}\langle \mathbf{d}(t)-\mathbf{h}(t,\Theta)|\mathbf{d}(t)-\mathbf{h}(t,\Theta)\rangle+C_0 \\
&= \ln\Lambda(\Theta)-\frac{1}{2}\langle \mathbf{d}|\mathbf{d}\rangle+C_0,
\end{aligned}
\end{equation}
where $C_0$ is a normalization constant. The central statistic is the log-likelihood ratio, $\ln\Lambda(\Theta)$:
\begin{equation}\label{eq:ll1}
	\ln\Lambda(\Theta)=\langle \mathbf{d}|\mathbf{h}(t,\Theta)\rangle-\frac{1}{2}\langle \mathbf{h}(t,\Theta)|\mathbf{h}(t,\Theta)\rangle.
\end{equation}
The inner product $\langle \cdot | \cdot \rangle$ weights the data and template by the noise properties. For discrete time-series data, it takes the form:
\begin{equation}
	\langle \mathbf{h}_1(t) | \mathbf{h}_2(t)\rangle = \mathbf{h}_1(t)\mathcal{C}^{-1}\mathbf{h}_2^{\intercal}(t),
\end{equation}
where $\mathcal{C}$ is the noise covariance matrix and $\intercal$ denotes the transpose.

A robust estimation of $\mathcal{C}$ is critical for the analysis. We construct it from the \ac{ACF} of the detector noise, which is derived from a multi-stage data conditioning process following the procedures in \citep{Wang:2023mst,Wang:2024liy}. 
First, the raw data is downsampled from $4096$ Hz to $1024$ Hz using a Butterworth filter, which is sufficient for the $~63$ Hz frequency of the fundamental mode. 
The data is then high-pass filtered at $20$ Hz with a Finite Impulse Response filter \citep{KhanFIR2020}. 
From this cleaned data, we estimate the one-sided \ac{PSD} using the Welch method \citep{1967D.Welch} with an \textbf{inverse spectrum truncation} algorithm. All procedures are implemented using the \textbf{PyCBC} package (version $2.7.2$) \citep{2012PhRvD..85l2006A}. Finally, the \ac{ACF} is obtained via an inverse Fourier transform of the \ac{PSD}, from which the Toeplitz covariance matrix $\mathcal{C}$ is constructed.

To ensure this \ac{TD} framework is reliable, the data processing procedure is subjected to extensive validation. The resulting \ac{ACF} not only passes a consistency test between full \ac{IMR} analyses in the time and frequency domains \cite{Wang:2024liy}, but has also been shown to produce consistent results for \ac{TD} ringdown analyses across different sampling rates \cite{Wang:2023mst}. These validations confirm that our time-domain noise model does not introduce biases, avoids artifacts from improper data handling such as flawed downsampling methods, and yields consistent results for \ac{TD} ringdown analyses across different sampling rates \citep{Wang:2023mst, Wang:2024liy}. 
For the final ringdown analysis, we use a $0.4$ s truncated segment of the validated $8$ s \ac{ACF}. 
This entire process of sampling the full parameter set $\Theta$ with the likelihood defined above is what we refer to as \ac{TTD}.

\subsection{The $\mathcal{F}$-statistic}
As an alternative to \ac{TTD}, we employ the $\fs$ method, a semi-analytical approach designed to enhance computational efficiency by reducing the dimensionality of the parameter space. The core principle is to separate the waveform parameters into two groups: a ``linear" set, which can be analytically maximized out of the likelihood, and a ``non-linear" set, which must still be explored with stochastic sampling.

For the time-domain ringdown model, this separation is achieved by recasting the physical amplitudes $A_{\ell mn}$ and phases $\phi_{\ell mn}$ of each \ac{QNM} into a pair of Cartesian amplitude parameters, $B^{\ell mn,1}$ and $B^{\ell mn,2}$. This reformulation allows the total signal in a detector, $\mathbf{h}(t)$, to be expressed as a linear combination of basis waveforms $\mathbf{g}_\mu(t)$:
$$\mathbf{h}(t) = B^\mu \mathbf{g}_\mu(t).$$
In this expression, the Einstein summation convention is used\footnote{In this work, we adopt the Einstein summation convention.}, the linear coefficients $B^\mu$ and corresponding basis waveforms $\mathbf{g}_\mu$ for each mode $(\ell, m, n)$ are defined as:
\begin{equation}
\begin{aligned}
B^{\ell mn,1} = &A_{\ell mn}\cos\phi_{\ell mn}, \\
B^{\ell mn,2} = &A_{\ell mn}\sin\phi_{\ell mn}, \\
\mathbf{g}_{\ell mn,1} = &\left[F^+\cos(2\pi f_{\ell mn}t)+F^{\times}\sin(2\pi f_{\ell mn}t)\right]\\
					   &\times {}_{-2}Y_{\ell m}(\iota,\delta)\exp(-t/\tau_{\ell mn}), \\
\mathbf{g}_{\ell mn,2} = &\left[-F^{+}\sin(2\pi f_{\ell mn}t)+F^{\times}\cos(2\pi f_{\ell mn}t)\right] \\
					   &\times {}_{-2}Y_{\ell m}(\iota,\delta)\exp(-t/\tau_{\ell mn}).
\end{aligned}
\end{equation}
The basis waveforms $\mathbf{g}_\mu$ depend on the set of non-linear parameters $\bm\theta = \{\text{RA}, \text{DEC}, t_c,\psi,\iota,M_f,\chi_f\}$, which form the reduced parameter space for sampling.

By substituting this reformulated waveform into the log-likelihood ratio from Eq.~\eqref{eq:ll1}, the expression becomes a quadratic function of the linear parameters $B^\mu$:
\begin{equation}\label{eq:ll2}
	\ln\Lambda(\Theta)=B^{\mu}\mathbf{s}_{\mu}(\bm\theta)-\frac{1}{2}B^{\mu}\mathbf{M}_{\mu\nu}(\bm\theta)B^{\nu}.
\end{equation}
Here, $\mathbf{s}_{\mu} \equiv \langle d|\mathbf{g}_{\mu}\rangle$ is the projection of the data onto the basis waveforms, and $\mathbf{M}_{\mu\nu} \equiv \langle \mathbf{g}_{\mu}|\mathbf{g}_{\nu}\rangle$ is the covariance matrix of these waveforms. For any given point $\bm\theta$ in the non-linear parameter space, this likelihood has a unique maximum with respect to $B^\mu$ that can be found analytically. Setting the partial derivative $\partial\ln\Lambda / \partial B^{\nu}$ to zero yields a linear system of equations whose solution is the maximum likelihood estimator for the linear parameters, $\hat{B}^{\mu}$:
\begin{equation}\label{eq:B}
\hat{B}^{\mu}=\mathbf{M}^{\mu\nu}\mathbf{s}_{\nu},
\end{equation}
where $\mathbf{M}^{\mu\nu}$ is the inverse of $\mathbf{M}_{\mu\nu}$. Notably, this estimator is independent of the prior on $B^\mu$.

Substituting this analytic solution back into Eq.~\eqref{eq:ll2} yields the profile likelihood. As shown in \citep{Wang:2024jlz, Wang:2024yhb}, this allows the log-likelihood ratio to be separated into two components:
\begin{equation}\label{eq:fs0}
\ln\Lambda(\Theta)=\frac{1}{2}\mathbf{s}_{\mu}\mathbf{M}^{\mu\nu}\mathbf{s}_{\nu}-\frac{1}{2}(B^{\mu}-\hat{B}^{\mu})\mathbf{M}_{\mu\nu}(B^{\nu}-\hat{B}^{\nu}).
\end{equation}
This rearranged form is powerful because it allows for a direct factorization of the posterior. According to the conjunction rule of probability, the joint posterior can be written as $P(\bm\theta,B^{\mu}|d)=P(\bm\theta|d)P(B^{\mu}|\bm\theta,d)$, which maps directly onto the two terms in Eq.~\eqref{eq:fs0}. The first term, which depends only on the non-linear parameters $\bm\theta$, defines the $\fs$:
\begin{align}\label{eq:fs1}
\ln\Lambda(\bm\theta) \equiv 2\mathcal{F} = \frac{1}{2}\mathbf{s}_{\mu}\mathbf{M}^{\mu\nu}\mathbf{s}_{\nu}.
\end{align}
This is the statistic used as the log-profile likelihood for the stochastic sampling of $\bm\theta$. The second term describes the conditional posterior for the linear parameters, $P(B^{\mu}|\bm\theta,d)$, which is a multivariate Gaussian centered at $\hat{B}^{\mu}$ with covariance $\mathbf{M}^{\mu\nu}$. For analyses involving multiple detectors, the network $\fs$ is simply the sum of the statistics from each detector.

\subsection{Parameter Estimation Strategy and Priors}\label{ssec:plp}
The use of the $\fs$ necessitates a two-step strategy for parameter estimation. The first step consists of the primary analysis, where we efficiently explore the reduced-dimension, non-linear parameter space of $\bm\theta$ using the profile likelihood (Eq.~\eqref{eq:fs1}) in a stochastic sampler. The second step involves a post-processing reconstruction of the posterior distributions for the ``linear" parameters ($B^\mu$) that were analytically maximized during the first step.

This reconstruction is possible because the conditional posterior for the linear parameters, $P(B^\mu | \boldsymbol{\theta}, \mathbf{d})$, is a known multivariate Gaussian for any given sample $\boldsymbol{\theta}_i$ from the primary analysis:
\begin{equation}
	P(B^\mu | \boldsymbol{\theta}_i, \mathbf{d}) \sim \mathcal{N}\left(\hat{B}^\mu(\boldsymbol{\theta}_i), \mathbf{M}^{\mu\nu}(\boldsymbol{\theta}_i)\right).
\end{equation}
The full marginal posterior, $P(B^\mu | \mathbf{d})$, is then constructed by marginalizing over the uncertainty in the non-linear parameters. We achieve this via composition: for each posterior sample $\boldsymbol{\theta}_i$, we draw a corresponding set of samples for $B^\mu$ from its conditional Gaussian distribution. The resulting collection of all $B^\mu$ samples represents the full posterior.

A crucial aspect of this procedure is the choice of prior. The reconstruction method described above most naturally operates with a simple, uniform sampling prior on the Cartesian amplitude parameters $B^\mu$. The relationship between the Cartesian priors and the physical priors on amplitude ($A_{\ell mn}$) and phase ($\phi_{\ell mn}$) is given by the Jacobian of the transformation:
\begin{equation}\label{eq:pb}
	p(B^{\ell mn,1},B^{\ell mn,2})=A_{\ell mn}p(A_{\ell mn},\phi_{\ell mn}).
\end{equation}
A uniform prior on $B^\mu$ corresponds to a desirable uniform prior on the phase, $\phi_{\ell mn} \in [0, 2\pi]$, but also implies an unphysical prior on the amplitude, $p(A_{\ell mn}) \propto 1/A_{\ell mn}$.

To obtain results with a physically motivated prior, we use importance resampling. This technique re-weights the samples drawn during the reconstruction to transform the posterior from the one generated under the unphysical sampling prior to the one corresponding to our desired target prior. The weight for each sample is the ratio of the target prior probability to the sampling prior probability:
\begin{equation}\label{eq:iw1}
w_i = \frac{p(B^{\mu}_i | \boldsymbol{\theta}_i, \otimes)}{p(B^{\mu}_i | \boldsymbol{\theta}_i,\ominus)},
\end{equation}
where $\ominus$ and $\otimes$ denote the uniform (sampling) and target priors on $B^{\mu}$, respectively.

For the final results presented in this work, we adopt a set of physically motivated target priors. For the linear parameters, we assume the amplitudes of all modes are distributed uniformly in the range $[0, 5\times 10^{-19}]$. For the non-linear parameters, we use flat priors within the following ranges: $M_f\in[100, 500]\,\Msun$, $\chi_f\in[0, 0.99]$, and $\cos\iota\in[-1,1]$. The phase parameter $\phi_{\ell mn}$ is also assumed to be uniform in $[0,2\pi]$. This complete set of priors is used for both the $\fs$ analysis (as the target prior) and the \ac{TTD} analysis to ensure a consistent comparison between the two methods.

\subsection{Evidence and Bayes Factor Calculation}
A primary advantage of Bayesian inference is the ability to perform model selection using the Bayes factor, $\mathcal{B}$, which is the ratio of the Bayesian evidence, $\mathcal{Z}$, for two competing models. While nested sampling algorithms directly compute the evidence for \ac{TTD} by integrating over the full parameter space $\Theta$, this is not the case for the $\fs$ method. The primary output of an $\fs$ analysis is a partial evidence, $\mathcal{Z}(\bm\theta)$, that only accounts for the stochastically sampled non-linear parameters $\bm\theta$.

To enable robust model comparison, we implement a novel formulation to reconstruct the full evidence, $\mathcal{Z}(\Theta)$, from the outputs of the $\fs$ run. The method correctly incorporates the contribution of the analytically maximized linear parameters $B^\mu$ under a chosen target prior (denoted by $\otimes$). The full evidence can be expressed as an expectation value:
\begin{align}
\mathcal{Z}(\Theta|\otimes) &= \int \mathcal{L}(\bm\theta, B^{\mu}) p(\bm\theta, B^{\mu} | \otimes) d\bm\theta dB^{\mu} \notag\\
&= \int \left[ \mathcal{L}(\bm\theta) f(\bm\theta) \right] p(\bm\theta) d\bm\theta.
\end{align}
This integral is approximated as an average over the $N(\bm\theta)$ posterior samples of the non-linear parameters:
\begin{equation}
\mathcal{Z}(\Theta|\otimes) \approx \frac{\mathcal{Z}(\bm\theta)}{N(\bm\theta)} \sum_{i=1}^{N(\bm\theta)} f(\bm\theta_i).\label{eq:z1}
\end{equation}
The weighting function, $f(\bm\theta_i)$, represents the evidence of the linear parameters for each non-linear sample $\bm\theta_i$. It is calculated via importance sampling, re-weighting from the simple uniform prior (denoted by $\ominus$) used in the reconstruction to the desired target prior ($\otimes$):
\begin{align}
f(\bm\theta) &= \int \Lambda(B^{\mu} | {\bm\theta}) p(B^{\mu} | {\bm\theta}, \otimes) dB^{\mu} \notag\\
&\approx \frac{\mathcal{Z}(B^{\mu}|{\bm\theta}, \ominus)}{N(B^{\mu}|{\bm\theta}, \ominus)} \sum_{j=1}^{N(B^{\mu}|{\bm\theta}, \ominus)} \frac{p(B^{\mu}_j | {\bm\theta}, \otimes)}{p(B^{\mu}_j | {\bm\theta}, \ominus)}.\label{eq:f1}
\end{align}
In this study, we draw $100\times N_0^2$ samples from the conditional posterior for this calculation, where $N_0$ is the number of \acp{QNM} in the model. With the full evidence $\mathcal{Z}_{\rm Fs}$ computed, we can calculate the Bayes factor $\mathcal{B}^{\rm w1}_{\rm w2} = \mathcal{Z}_{\rm w1} / \mathcal{Z}_{\rm w2}$ to quantitatively compare different ringdown waveform models.

\subsection{Sampler Configuration}
All Bayesian inference analyses were conducted using the {\sc Bilby} package~\citep[v2.4.0;][]{Ashton:2018jfp}, with {\sc Dynesty} \citep[v2.1.5;][]{Romero-Shaw:2020owr} as the nested sampling engine. We configured the sampler with $1000$ live points and used the ``rwalk" method for the primary sampling. The random walk was configured with $20$ autocorrelation lengths, and proposals were drawn using a mix of ``diff" and ``volumetric" methods. New points were accepted using the ``live-multi" selection scheme. This set of configurations follows the standard practices outlined in the {\sc Bilby} documentation to ensure robust and efficient exploration of the parameter space. To reduce computational time, all analyses were parallelized across $20$ CPU threads.

\section{Two-Mode Combination Results}\label{A2:2m}

\begin{figure*}
\centering
\begin{subfigure}[b]{0.48\linewidth}
\centering
\includegraphics[width=\textwidth,height=8cm]{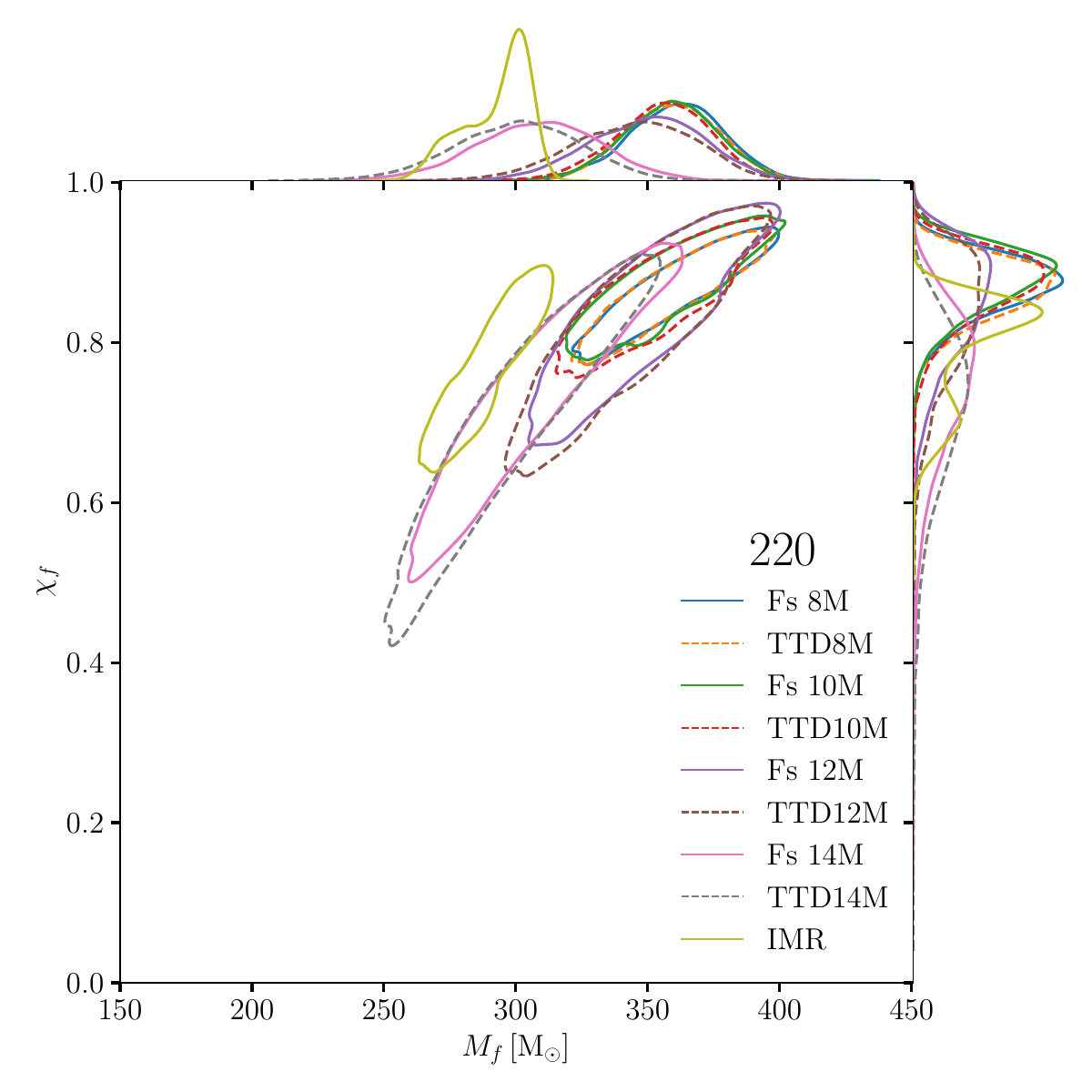}
\end{subfigure}%
\begin{subfigure}[b]{0.48\linewidth}
\centering
\includegraphics[width=\textwidth,height=8cm]{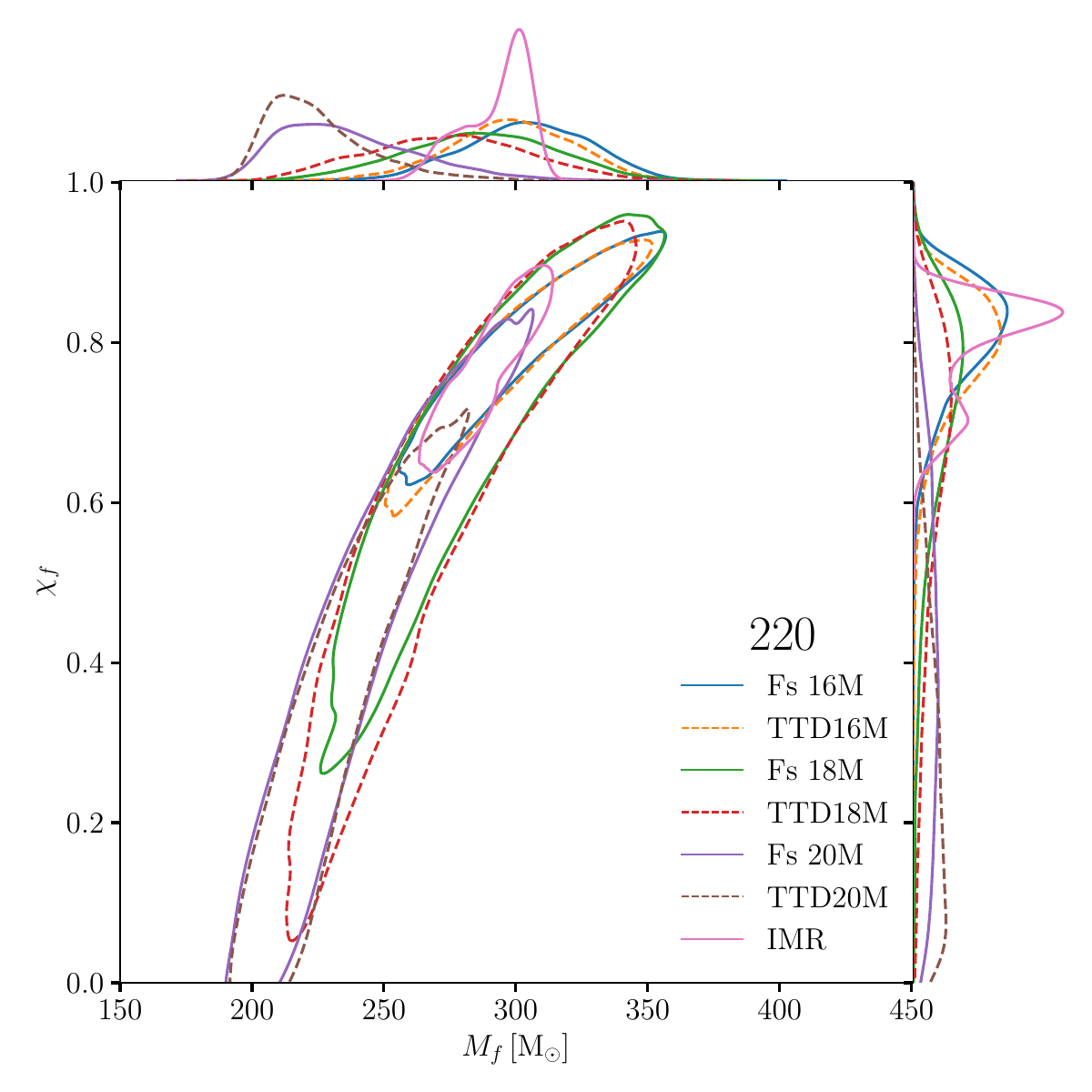}
\end{subfigure}\\
\begin{subfigure}[b]{0.48\linewidth}
\centering
\includegraphics[width=\textwidth,height=8cm]{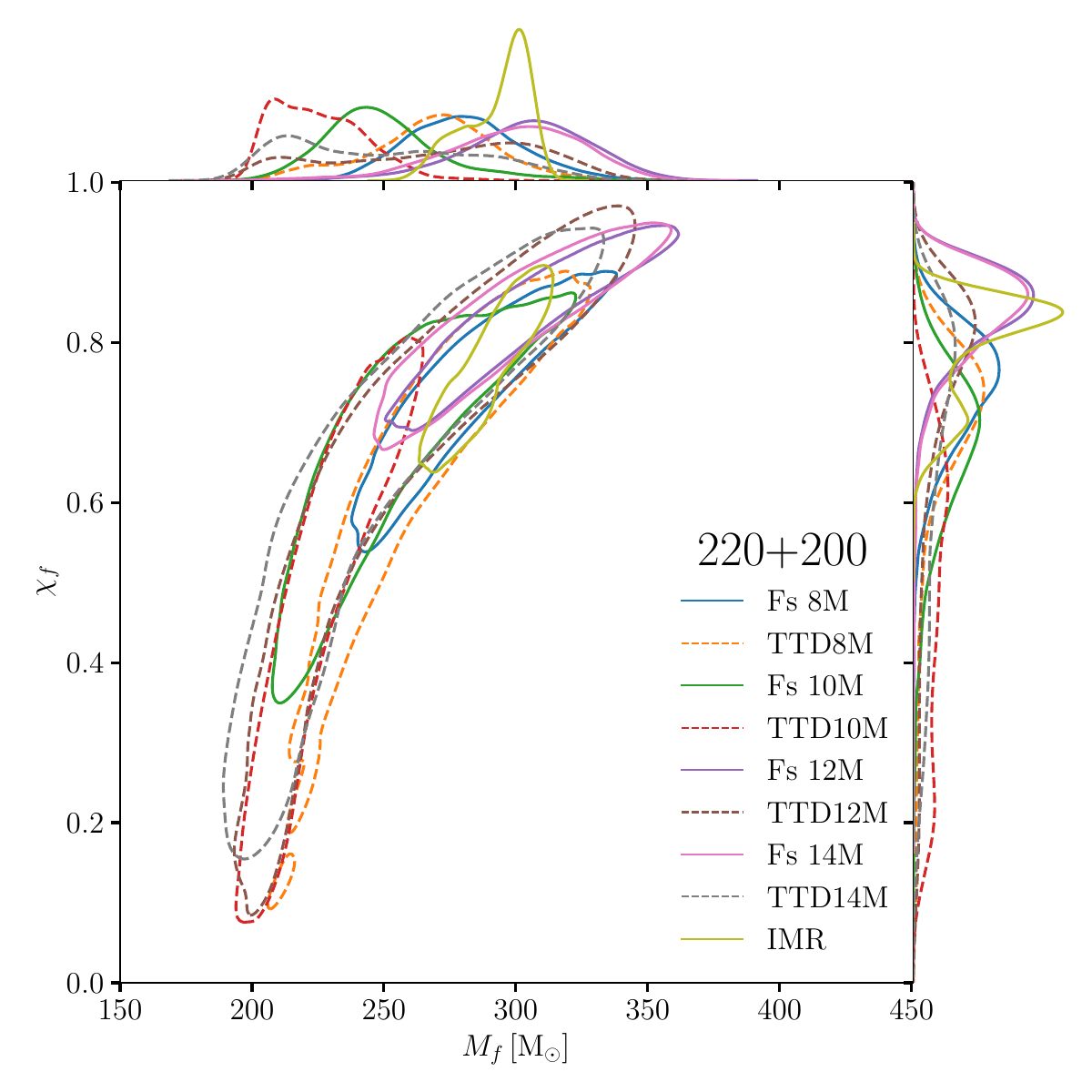}
\end{subfigure}%
\begin{subfigure}[b]{0.48\linewidth}
\centering
\includegraphics[width=\textwidth,height=8cm]{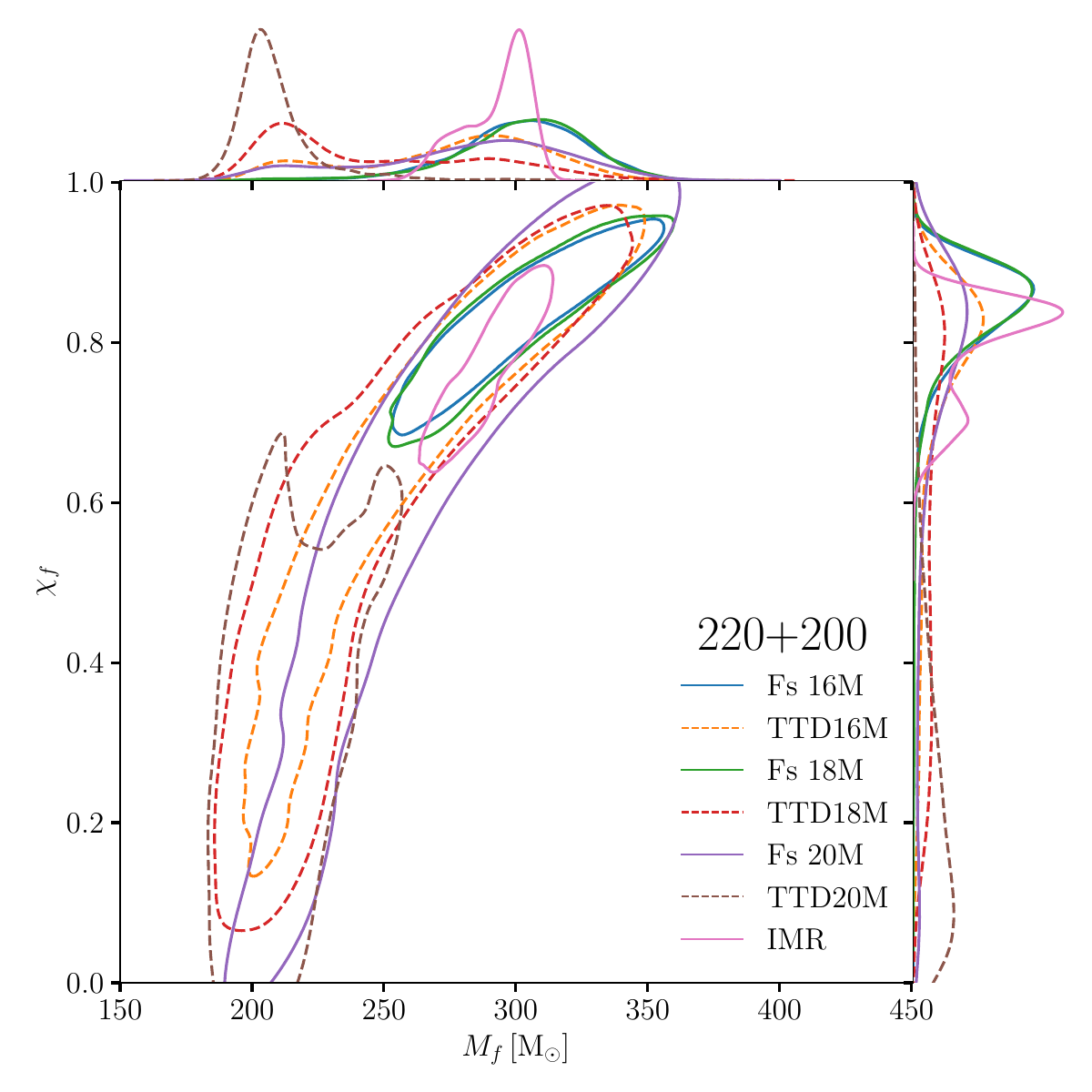}
\end{subfigure}%
\caption{
The posterior distributions of the redshifted final mass $M_f$, and the final spin $\chi_f$, of the GW231123 remnant were obtained utilizing the \ac{TTD} and the $\fs$ method. 
Results derived from fundamental-mode-only analyses are displayed on the left panel, while those incorporating the $(\ell |m|n)=(200)$ mode are presented on the right panel. 
The top two panels show results based on the fundamental mode only analyses with different start times, i.e., from $\Delta t=8\,M$ to $14\,M$ (left panel) and from $\Delta t=16\,M$ to $20\,M$ (right panel).
The bottom two panels show results based on the combination $220+201$ with different start times, i.e., from $\Delta t=8\,M$ to $14\,M$  (left panel) and from $\Delta t=16\,M$ to $20\,M$ (right panel).
The full \ac{IMR} analysis result based on the \textbf{NRSur7dq4} waveform model \citep{LIGOScientific:2025rsn} is indicated by the yellow contour.
Ringdown analysis results using the \ac{TTD} method and the $\fs$ method are indicated by dashed and solid contours at $90\%$ credible level, respectively.
Additionally, the marginal posterior distributions for both $M_f$ and $\chi_f$ are shown in their respective top and right panels.
}\label{fig:fmfs1_m2}
\end{figure*}

\begin{figure*}
\centering
\begin{subfigure}[b]{0.48\linewidth}
\centering
\includegraphics[width=\textwidth,height=8cm]{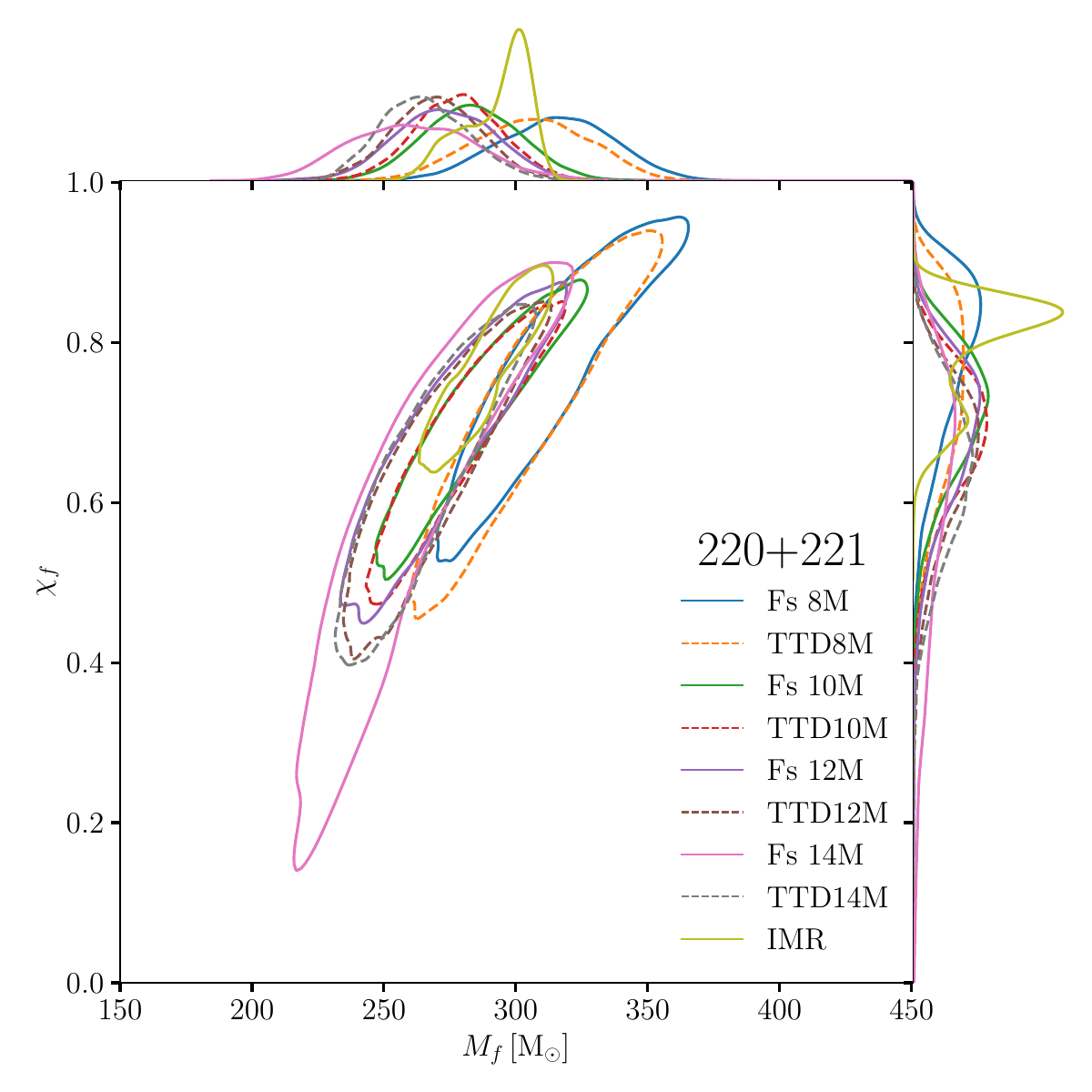}
\end{subfigure}%
\begin{subfigure}[b]{0.48\linewidth}
\centering
\includegraphics[width=\textwidth,height=8cm]{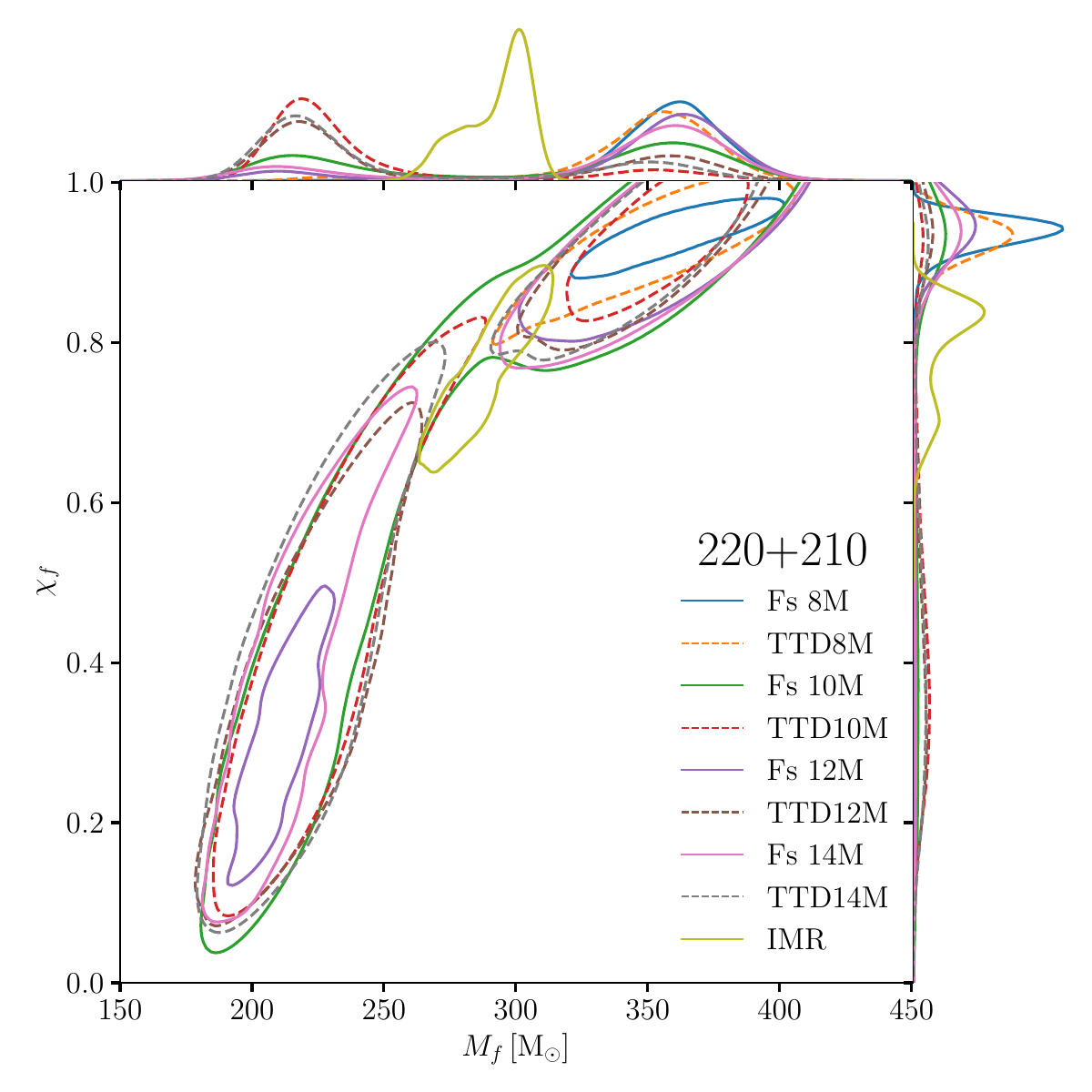}
\end{subfigure}\\
\begin{subfigure}[b]{0.48\linewidth}
\centering
\includegraphics[width=\textwidth,height=8cm]{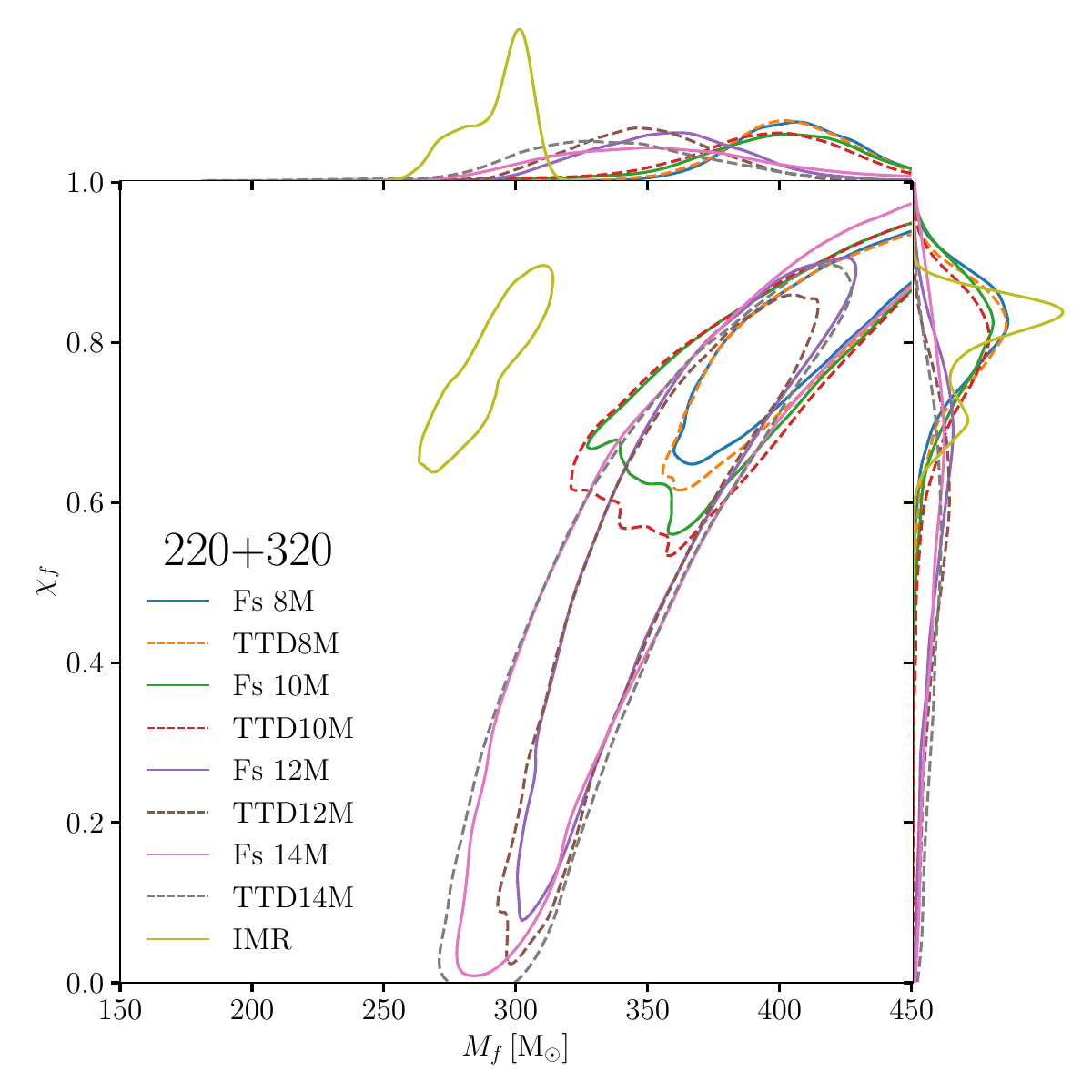}
\end{subfigure}%
\begin{subfigure}[b]{0.48\linewidth}
\centering
\includegraphics[width=\textwidth,height=8cm]{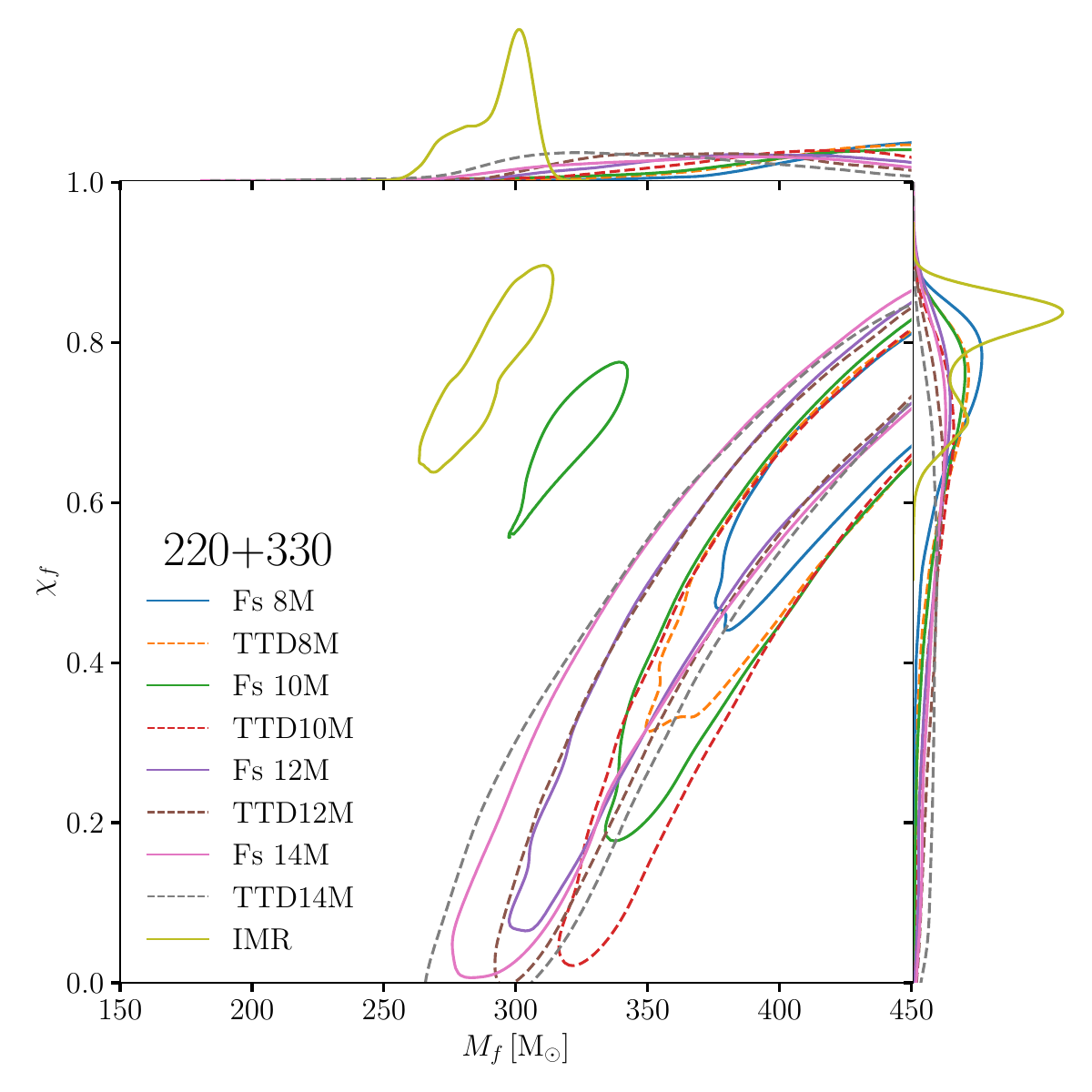}
\end{subfigure}\\
\begin{subfigure}[b]{0.48\linewidth}
\centering
\includegraphics[width=\textwidth,height=8cm]{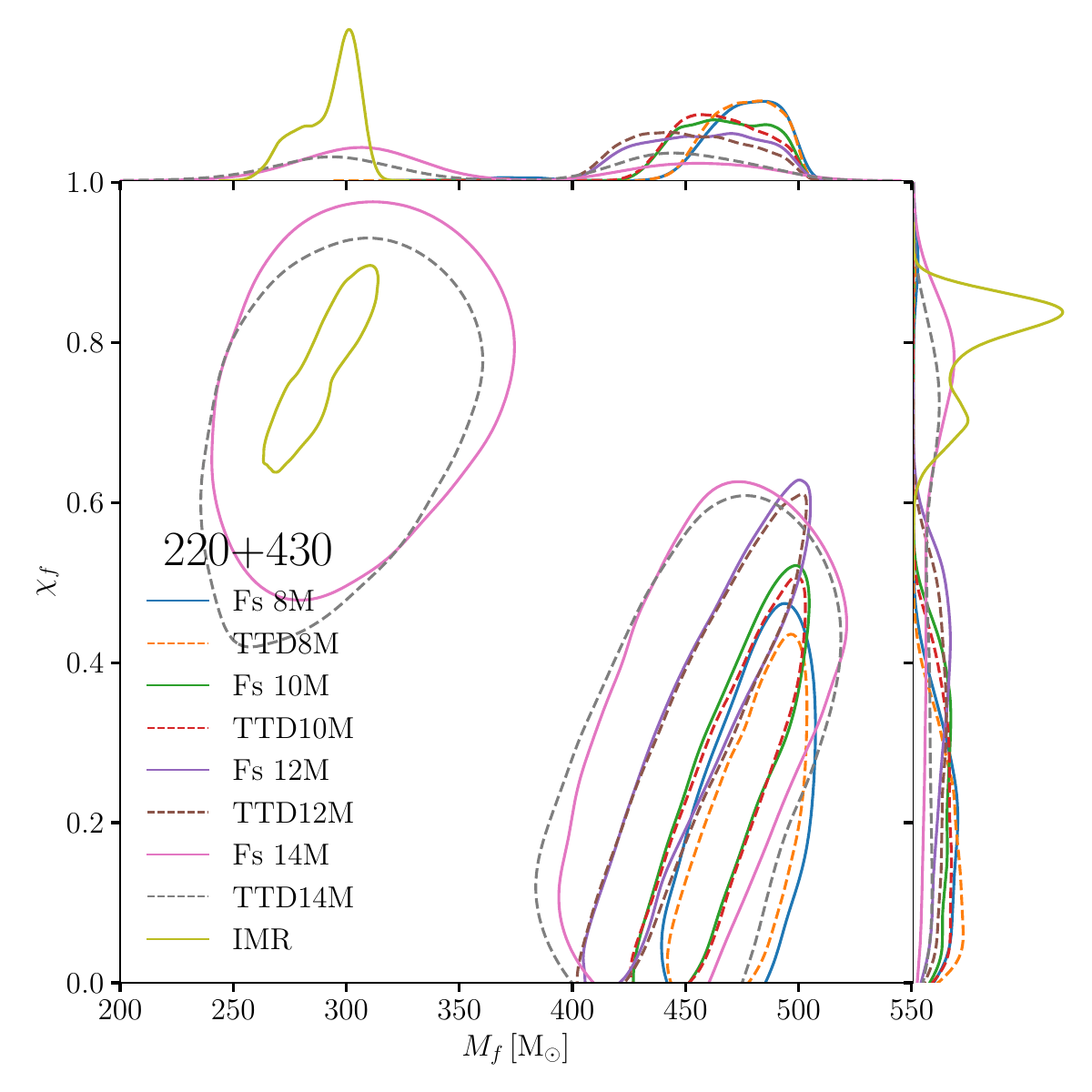}
\end{subfigure}%
\begin{subfigure}[b]{0.48\linewidth}
\centering
\includegraphics[width=\textwidth,height=8cm]{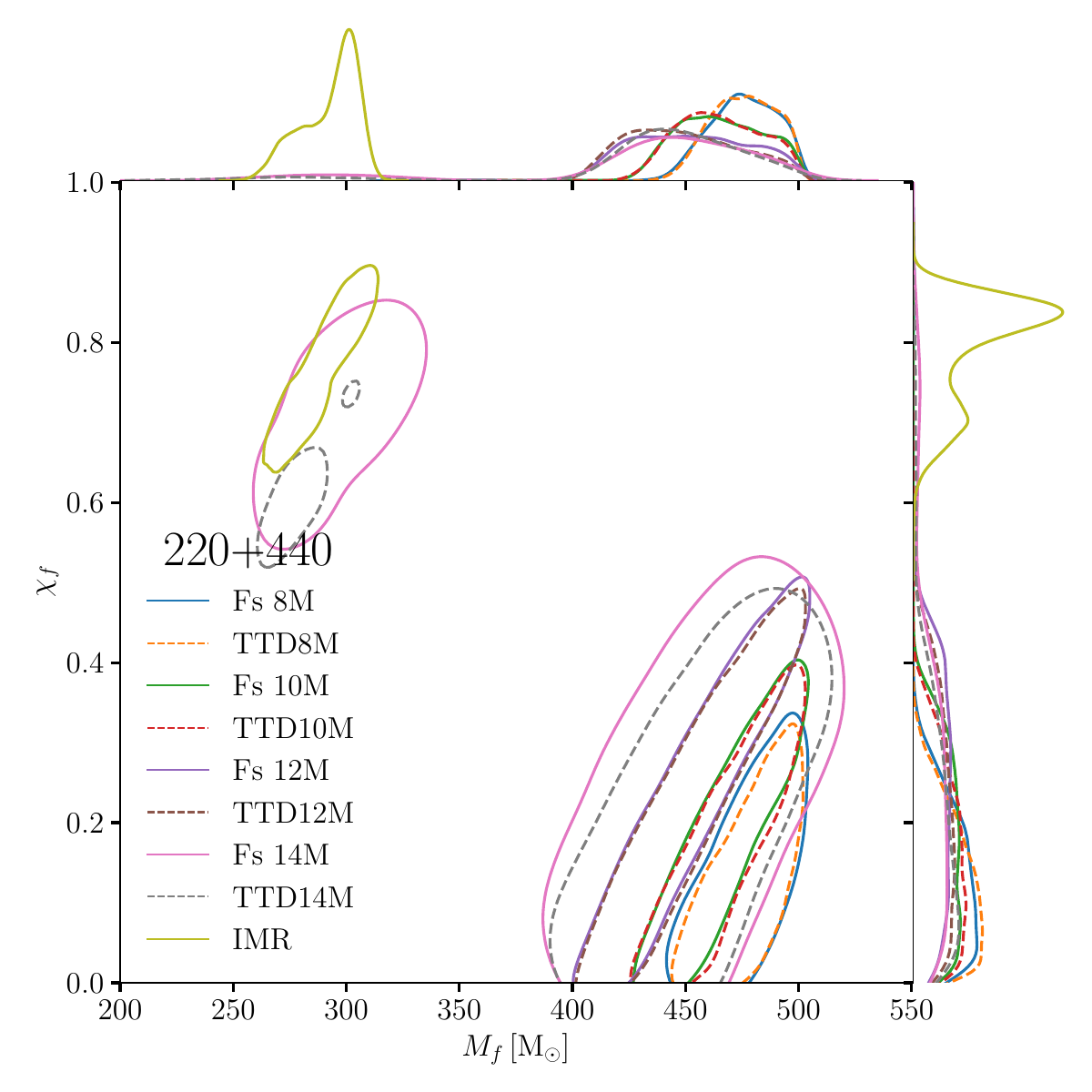}
\end{subfigure}\\
\caption{
Similar to Fig.~\ref{fig:fmfs1_m2}.
The top two panels show results based on the combinations $220+221$ (left panel) and $220+210$ (right panel) with different start times, ranging from $\Delta t=8\,M$ to $14\,M$.
The middle two panels show results based on the combinations $220+320$ (left panel) and $220+330$ (right panel).
The bottom two panels show results based on the combinations $220+430$ (left panel) and $220+440$ (right panel).
}\label{fig:fmfs2_m2}
\end{figure*}

This section provides a more detailed comparison of the posterior distributions for the two-mode ringdown analyses, complementing the discussion in the main text. Figs. \ref{fig:fmfs1_m2} and \ref{fig:fmfs2_m2} show the results for the fundamental-mode-only ($220$) model and for models combining the fundamental mode with a second mode from the set $\{221, 200, 210, 320, 330, 430, 440\}$. For the most relevant cases ($220$ only and $220+200$), we show results for start times ranging from $\Delta t = 8\,M$ to $20\,M$. For the other, less-supported combinations, we only show results up to $\Delta t = 14\,M$, as their Bayes factors decrease significantly at later times.

The fundamental-mode-only analysis struggles to produce posteriors that are consistent with the full \ac{IMR} result until late start times ($\Delta t \ge 16\,M$). At much later times, the diminished \ac{SNR} leads to very broad and uninformative constraints, particularly for the final spin. In contrast, the $220+200$ combination yields posteriors that are in significant agreement with the \ac{IMR} analysis over a wide range of start times. Notably, the bimodal features in the final mass and spin posteriors that appear in the \ac{TTD} analysis (consistent with findings in \citet{LIGOScientific:2025rsn}) are substantially suppressed in the results from the $\fs$ method, especially for start times from $\Delta t = 12\,M$ to $18\,M$.

\begin{figure*}
\centering
\begin{subfigure}[b]{0.75\linewidth}
\centering
\includegraphics[width=\textwidth,height=8cm]{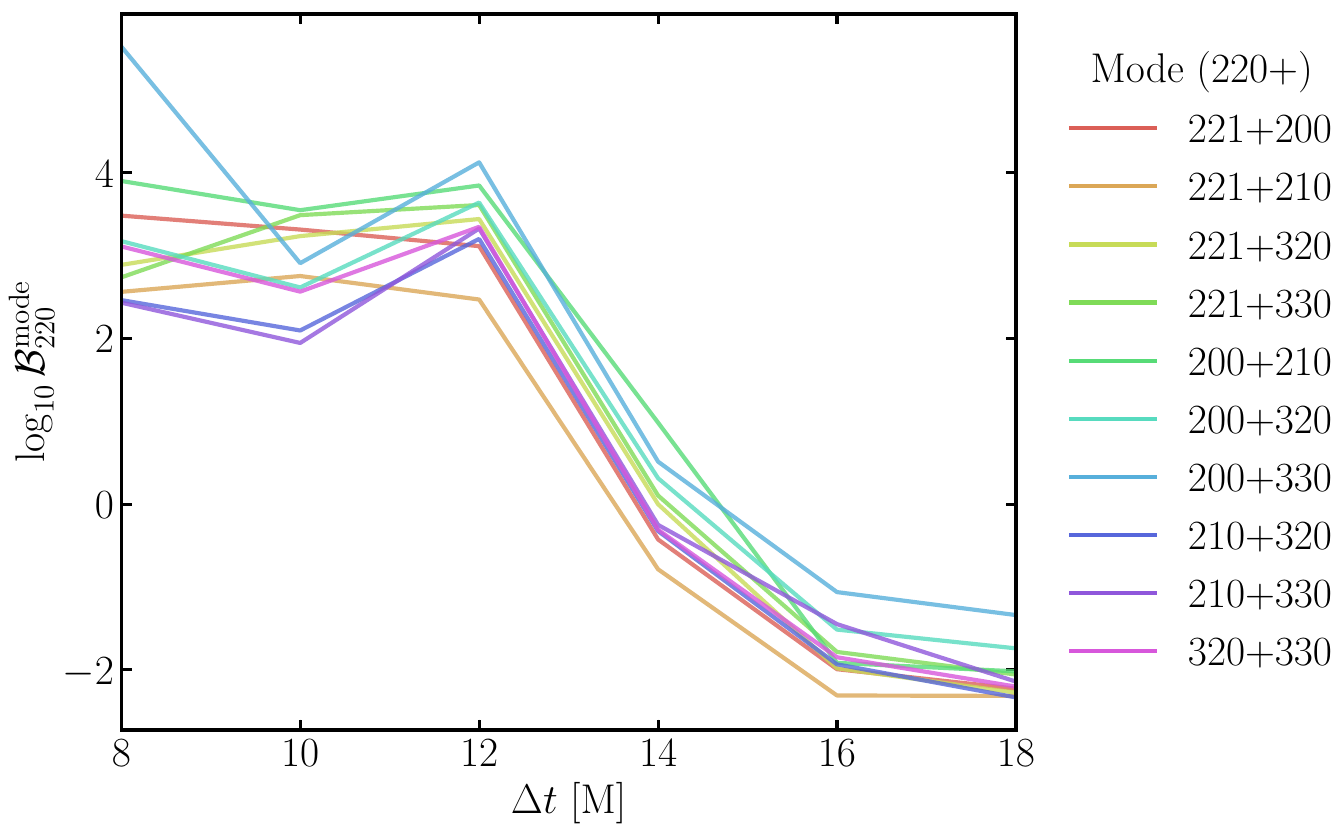}
\end{subfigure}\\
\begin{subfigure}[b]{0.75\linewidth}
\centering
\includegraphics[width=\textwidth,height=8cm]{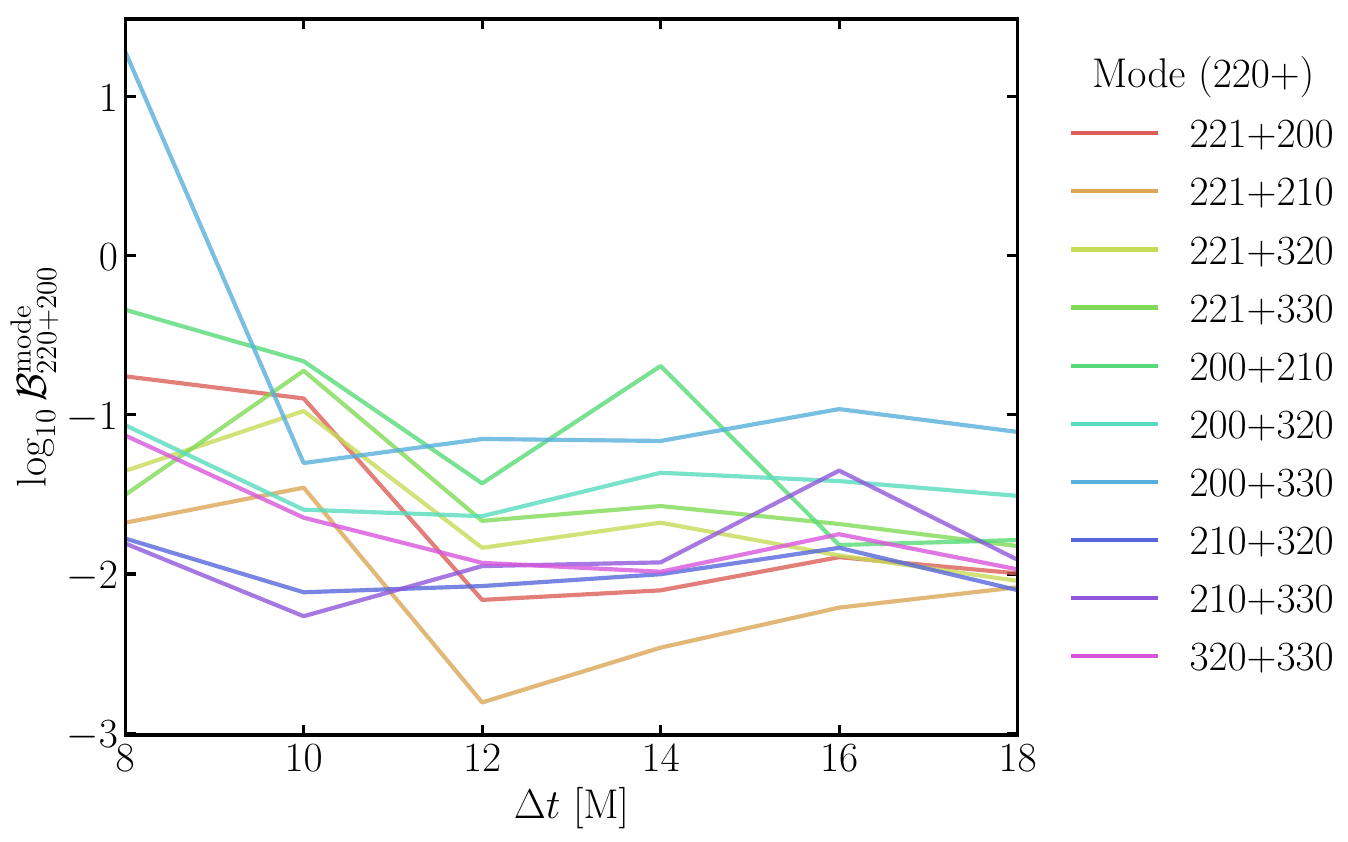}
\end{subfigure}\\
\caption{
Bayesian evidence for three \acp{QNM} combinations in the ringdown analysis of GW231123, quantified by the base-$10$ logarithm of the Bayes factor ($\log_{10} \mathcal{B}$) relative to the fundamental mode (top panel) and the combination $220+200$ (bottom panel) at each starting time. 
The horizontal axis shows the time delay ($\Delta t$) between the start of the ringdown analysis and the polarization peak. 
Results are based on the $\fs$ method.
}\label{fig:bfs_m3}
\end{figure*}

For the other two-mode combinations, the $220+221$ case is the only one that produces posteriors marginally consistent with the \ac{IMR} analysis (for $\Delta t$ from $10\,M$ to $14\,M$). However, its Bayes factor is much smaller than that of the $220+200$ case. The remaining combinations do not produce meaningful posterior distributions that overlap with the \ac{IMR} results and are not supported by the Bayesian evidence.

\section{Three-Mode Combination Search}\label{A3:3m}

\begin{figure}
\centering
\includegraphics[width=0.5\textwidth,height=8cm]{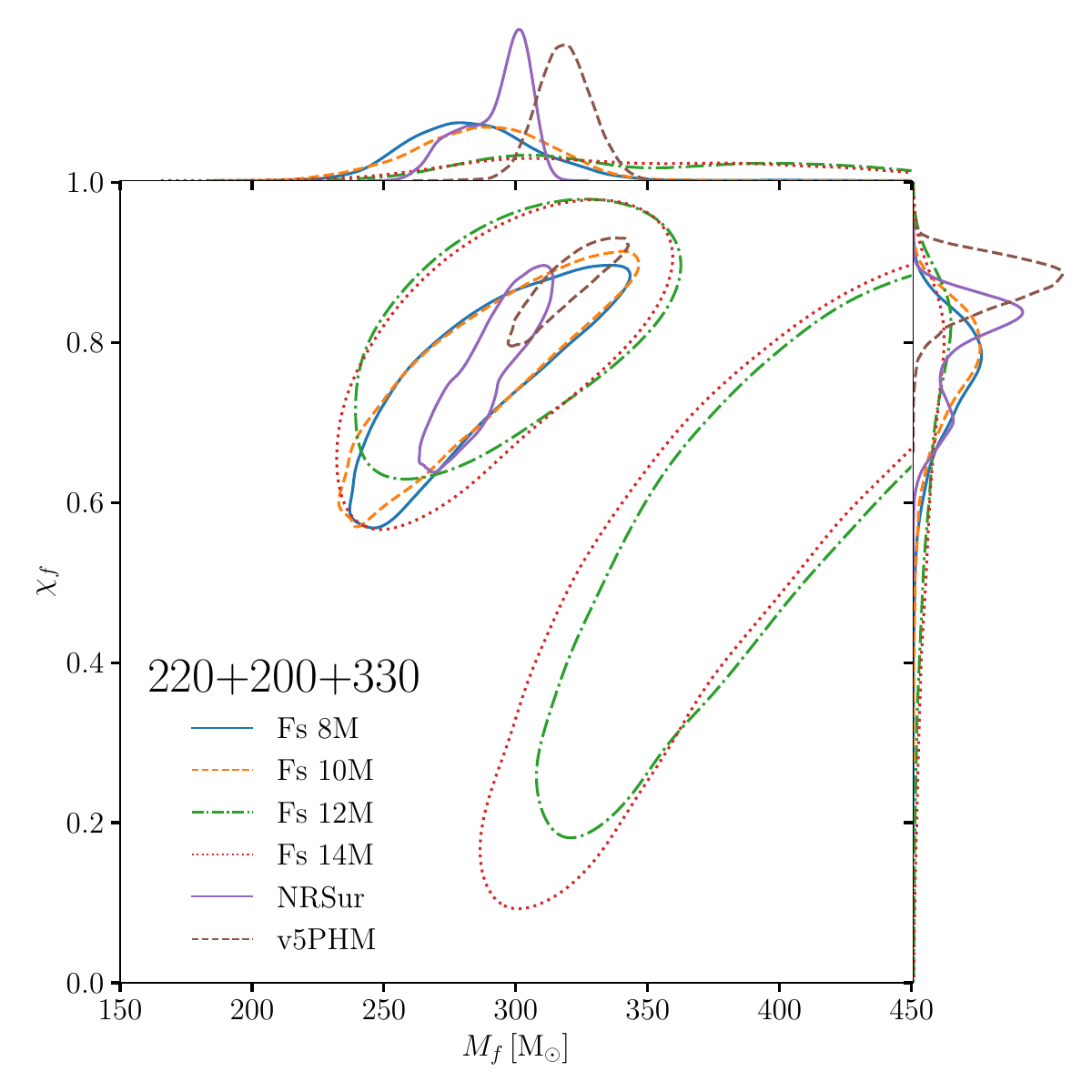}
\caption{
Constraints on the redshifted final mass (\(M_f\)) and final spin (\(\chi_f\)) from analyzing the GW231123 ringdown signal using the $\fs$ method (labeled ``Fs") with the $220+200+330$ mode combination. The analysis is performed at different start times (\(\Delta t = 8M\) to \(14M\)) relative to the polarization peak. Results are compared with those obtained from full \ac{IMR} analyses using the \textbf{NRSur7dq4} (labeled ``NRSur") and \textbf{SEOBNRv5PHM} (labeled ``v5PHM") waveform models. The central panel shows the $90\%$ credible regions (contours) for \(M_f\) and \(\chi_f\), while the top and right panels display the marginalized posterior distributions for final mass and final spin, respectively. 
}\label{fig:mfsf3}
\end{figure}

We further investigate the possibility of a third detectable \ac{QNM} in the signal. We performed analyses using three-mode combinations, consisting of the fundamental mode plus two additional modes selected from $\{221, 200, 210, 320, 330\}$. Fig.~\ref{fig:bfs_m3} shows the Bayes factors for the ten resulting combinations.

When compared to the fundamental-mode-only case, the $220+200+330$ combination yields the strongest evidence among the three-mode models, with its support peaking at a very early start time of $\Delta t=8\,M$. While the evidence for this three-mode model at $\Delta t=8\,M$ is stronger than that for the $220+200$ model at the same early time, it is not as strong as the overall maximum evidence found for the $220+200$ model at its optimal start time of $\Delta t=12\,M$.
 
Furthermore, Fig.~\ref{fig:mfsf3} shows that while this three-mode analysis yields remnant mass–spin posteriors that coincide with IMR results for early start times, a marked bimodality emerges as the start time increases (an effect absent in the $220+200$ analysis), with a secondary peak that is significantly offset from IMR-inferred values and overlaps the posteriors of the $220+330$ combination.

\end{document}